\documentclass{emulateapj}

\usepackage{fancyhdr}
\usepackage{color}
\usepackage[disable]{todonotes}

\newcommand{\msun}{\mbox{$M_\odot$}}
\newcommand{\kms}{\mbox{km s$^{-1}$}}
\newcommand{\dg}{\mbox{$^\circ$}}
\newcommand{\as}{\mbox{$^{\prime\prime}$}}
\def\farcsec{\hbox{$.\!\!^{\prime\prime}$}}
\newcommand{\twco}{\mbox{$^{12}\textrm{CO}$}}
\newcommand{\thco}{\mbox{$^{13}\textrm{CO}$}}

\def\vcloud{8\ \kms}
\newcommand{\tex}{\mbox{$T_{\textrm{ex}}$}}

\slugcomment{Received 2014 August 27; accepted 2015 February 10; published 2015 in the Astrophysical Journal.}

\shorttitle{Outflows in Serpens South}
\shortauthors{Plunkett et al.}

\begin{document}

\title{ASSESSING MOLECULAR OUTFLOWS AND TURBULENCE IN THE PROTOSTELLAR CLUSTER SERPENS SOUTH}

\author{Adele L. Plunkett\altaffilmark{1}, H\'{e}ctor G. Arce}
\affil{Department of Astronomy, Yale University,
    P.O. Box 208101, New Haven CT 06520, USA}

\author{Stuartt A. Corder }
\affil{Joint ALMA Observatory, Av. Alonso de C\'{o}rdova 3107, Vitacura, Santiago, Chile}

\author{Michael M. Dunham}
\affil{Harvard-Smithsonian Center for Astrophysics, 60 Garden Street, MS 78, Cambridge, MA 02138, USA}

\author{Guido Garay, Diego Mardones}
\affil{Departamento de Astronom\'{i}a, Universidad de Chile, Casilla 36-D, Santiago, Chile}

\altaffiltext{1}{NSF Graduate Research Fellow}

\begin{abstract}
Molecular outflows driven by protostellar cluster members likely impact their surroundings and contribute to turbulence, affecting subsequent star formation.  The very young Serpens South cluster  consists of a particularly high density and fraction of protostars, yielding a relevant case study for protostellar outflows and their impact on the cluster environment.  We combined CO $J=1-0$\ observations of this region using the Combined Array for Research in Millimeter-wave Astronomy (CARMA) and the Institut de Radioastronomie Millim\'{e}trique (IRAM) 30 m single dish telescope.  The combined map allows us to probe CO outflows within the central, most active region at size scales of 0.01 pc to 0.8 pc.  We account for effects of line opacity and excitation temperature variations by incorporating $^{12}$CO and $^{13}$CO data for the $J=1-0$\ and $J=3-2$\ transitions (using Atacama Pathfinder Experiment and Caltech Submillimeter Observatory observations for the higher CO transitions), and we calculate mass, momentum, and energy of the molecular outflows in this region.  The outflow mass loss rate, force, and luminosity, compared with diagnostics of turbulence and gravity, suggest that outflows drive a sufficient amount of energy to sustain turbulence, but not enough energy to substantially counter the gravitational potential energy and disrupt the clump.  Further, we compare Serpens South with the slightly more evolved cluster NGC 1333, and we propose an empirical scenario for outflow-cluster interaction at different evolutionary stages.
\end{abstract}


\section{Introduction}

\subsection{Molecular outflows and protostellar clusters}

Outflows and jets occur during the early stages of protostellar evolution, and they facilitate the accretion of material onto the protostar by carrying away excess angular momentum \citep{Bac96}.  A molecular outflow consists of ambient gas that is entrained by a protostellar jet, expected to occur during the Class 0 and I stages of star formation.  Class II sources also power jets and winds and may drive outflows, however molecular outflows are less prominent as a source evolves and the surrounding molecular gas becomes depleted.  Observationally, outflows serve as signposts that emanate from their driving protostellar sources and indicate that star formation is ongoing in a particular region.  Further, the characteristics and morphology of outflows may indicate the evolution of the forming protostar and constrain the gas entrainment mechanism \citep{Arc07}. 

Carbon monoxide is the most common probe of molecular outflows, because after H$_2$ it is the most prevalent molecule in molecular clouds and its low-level rotational states are easily excited in these environments \citep[e.g.,][]{Wil09}.  The CO molecular outflow components are generally characterized as standard high-velocity (SHV) with velocities of a few to about 20 \kms, and extremely high velocity (EHV) with even higher velocities, typically up to $\sim100$ \kms\ \citep[see][]{Bac96}.  Typically the SHV outflows are more massive than EHV, and in this paper we conduct a census of mostly SHV components within the protostellar cluster Serpens South in order to determine the impact of outflows on the surrounding cloud environment.

Outflows may reach distances up to several parsecs, affecting the surrounding protostellar environment corresponding to these scales, including the core and cloud.  Since outflows sweep up ambient gas, they influence the intimate relation between a protostar and its surroundings.   Considering that the majority of clustered star formation occurs within clusters of 100 or more members, and the size scales of clusters are typically less than a few parsecs \citep{Lad03}, there is ample opportunity for outflows to encounter other nearby outflows and their respective protostars, as well as potentially disrupt the surrounding cloud environment. 

It remains a challenge to quantify the cumulative impact of protostellar outflows on molecular clouds and subsequent star formation, but evidence of outflow-cloud interaction has been observed in a variety of star-forming regions \citep[e.g.,][]{Arc10, Nak11, Nak11Oph, Dua12}.  Specifically, outflows interact with the ambient cluster gas \citep{Ful02, Arc06}, inject momentum and energy into the surrounding cloud \citep{Arc07, Swi08}, and they may feed turbulent motions that affect ongoing star formation \citep[see also][]{Fra14}.  Models of clustered star formation must incorporate outflows in order to drive turbulence and produce outcomes close to what is observed in star-forming regions, including star formation rates (SFRs) and efficiencies \citep[e.g.,][]{Li06,Nak07,Car09,Wan10,Nak11Sim,Kru14}.  The cumulative effect of outflow activity, particularly in clustered star-forming regions, can be constrained by observations of clusters.

In the work presented here, we study outflow characteristics and assess their impact within Serpens South, which is the second case study in our survey of clustered star-forming regions ranging in evolutionary stage, beginning with NGC 1333 \citep{Plu13}.  NGC 1333 is a prototypical region that has been well-studied, while Serpens South is more recently discovered \citep{Gut08}, more massive and with a higher fraction of protostars in their earliest stages compared with NGC 1333. We describe Serpens South in this context in the following sub-section.  The goals of the current paper are: (a) measure the outflow activity in Serpens South, (b) investigate the sustained impact of outflows compared with turbulence and gravity, and (c) establish an empirical scenario of cluster evolution based on the comparison between Serpens South and NGC 1333.

\subsection{Description of the Region: Serpens South} \label{sec:SerpS}

The protostellar cluster Serpens South merits study for its youth, active star formation and corresponding outflows, the concentration of young stellar objects (YSOs) along a dense filamentary structure, and its relative proximity (429 pc, as discussed in the following paragraph). Serpens South was discovered by \citet{Gut08} in imaging of the Serpens-Aquila Rift with \textit{Spitzer} IRAC, in which its dark filamentary structure appeared as absorption in the 8 $\mu$m band.  Subsequent studies have confirmed the filamentary structure of this cloud using dust continuum maps \citep{And10,Mau11} and molecular line emission maps \citep{Kir13,Tan13,Fer14}. 

Serpens South is one of only a few active star-forming regions known within a distance $\lesssim$ 1 kpc away.  However, various distances have been adopted throughout the literature.  Serpens South and the Serpens Main cluster (3\dg\ to the north) were found to have similar local standard of rest velocities ($v_{LSR}$), in addition to their close proximity on the plane of the sky, and therefore the distance to Serpens South is based on the distance determined for Serpens Main \citep{Gut08}.  The majority of works since the discovery of Serpens South by \citet{Gut08} have assumed a distance to Serpens South of $260\pm37$ pc, which is the distance determined using photometry of Serpens Main by \citet{Str96} and consistent with the distance determined for the Serpens-Aquila Rift by \citet{Str03}.  Most recently, \citet{Dzi11} used Very Long Baseline Array trigonometric parallax measurements to determine a distance of $429\pm2$\ pc to the YSO EC 95 associated with Serpens Main, updated from the previous measurement of 415 pc by \citet{Dzi10}.  Throughout this paper, we assume that Serpens South is part of the same complex as Serpens Main, and we adopt the distance of $429\pm2$\ pc from \citet{Dzi11}.

Outflows in this region were observed by \citet{Nak11} at an angular resolution of $22\arcsec$, and presented in the context of the filamentary 1.1 mm dust continuum image by ASTE. Near-IR observations by \citet{Tei12} show corresponding molecular hydrogen emission-line objects. Polarization observations by \citet{Sug11} reveal the importance of the magnetic field in forming the main filament because the magnetic field is well-ordered and oriented perpendicular to this filament.  

The classification of candidate YSOs in Serpens South was first done by \citet{Gut08} following the data reduction pipeline presented by \citet{Eva07}, which was developed for the \textit{Spitzer} c2d and Gould Belt projects, incorporating 2MASS and \textit{Spitzer} $1.25-70$\ $\mu$m photometry.  Within the $14\arcmin\times10\arcmin$ region that they studied, \citet{Gut08} reported 54 Class I protostars (including flat-spectrum sources) and 37 Class II YSOs.  Hence, about 60\% of the YSOs are protostars, but at the cluster center this fraction rises to about 80\%, which is where we focus our study (see also Section \ref{sec:compare}).  \citet{Gut08} reported a mean YSO surface density of 430 pc$^{-2}$ at the distance of $260$ pc, which corresponds to $\sim160$\ pc$^{-2}$ at the farther distance of $429$ pc that we assume here.  The median projected distance between nearest-neighbor YSOs in this region is 13\farcs2 \citep[see][for discussion]{Gut08}, or 0.03 pc at a distance of 429 pc.  This nearest-neighbor spacing is less than the typical extent of outflows, and therefore this region provides an ideal case study where outflows likely interact with one another, the surrounding protostars and their cluster environment.

The \textit{Herschel} Gould Belt Survey \citep{And10,Bon10} revealed several of the youngest Class 0 members of Serpens South, further suggesting the overall youth of the region.  \citet{Mau11} suggest that the high number of protostars compared with the total number of YSOs in Serpens South indicates that this cluster experienced a recent burst of star formation, particularly along a filamentary structure spanning from the southeast to northwest.  They used complementary 1.2 mm continuum maps obtained with the MAx-Planck Millimeter BOlometer array (MAMBO) on the Institut de Radioastronomie Millim\'{e}trique (IRAM) 30 m telescope, along with \textit{Herschel} and \textit{Spitzer} maps, to perform a source extraction and classification over wavelengths from 2 $\mu$m to 1.2 mm.  They determined that this cluster has formed 9 Class 0, 48 Class I, and 37 Class II objects within an area of $15\arcmin \times 25\arcmin$ that has a mass of $\sim610\ \msun$.  They suggest that a high SFR and a relatively low star formation efficiency are evidence of the early evolutionary stage of this region.

Here we present the highest resolution observations published to date of millimeter-wavelength molecular line and continuum emission observations made toward the most active central region of Serpens South.  Specifically, we observed several isotopologues and energy level transitions of carbon monoxide, tracing the SHV outflow components. \twco\ traces cool outflow emission, a probe of swept up molecular gas.  \thco\ is less optically thick, and was used to obtain reliable estimates of outflow mass.  Together, observations of \twco\ and \thco\ provide a comprehensive view of the molecular gas kinematics in the region.  Continuum emission was simultaneously observed to detect dense dust envelopes around protostars.  Some of these protostars, along with the protostars observed in the previously mentioned studies, may be responsible for driving the outflows detected in line emission.

\section{Observations and Data} \label{sec:obs}

\begin{figure}[!ht]
\includegraphics[width=1.1\linewidth]{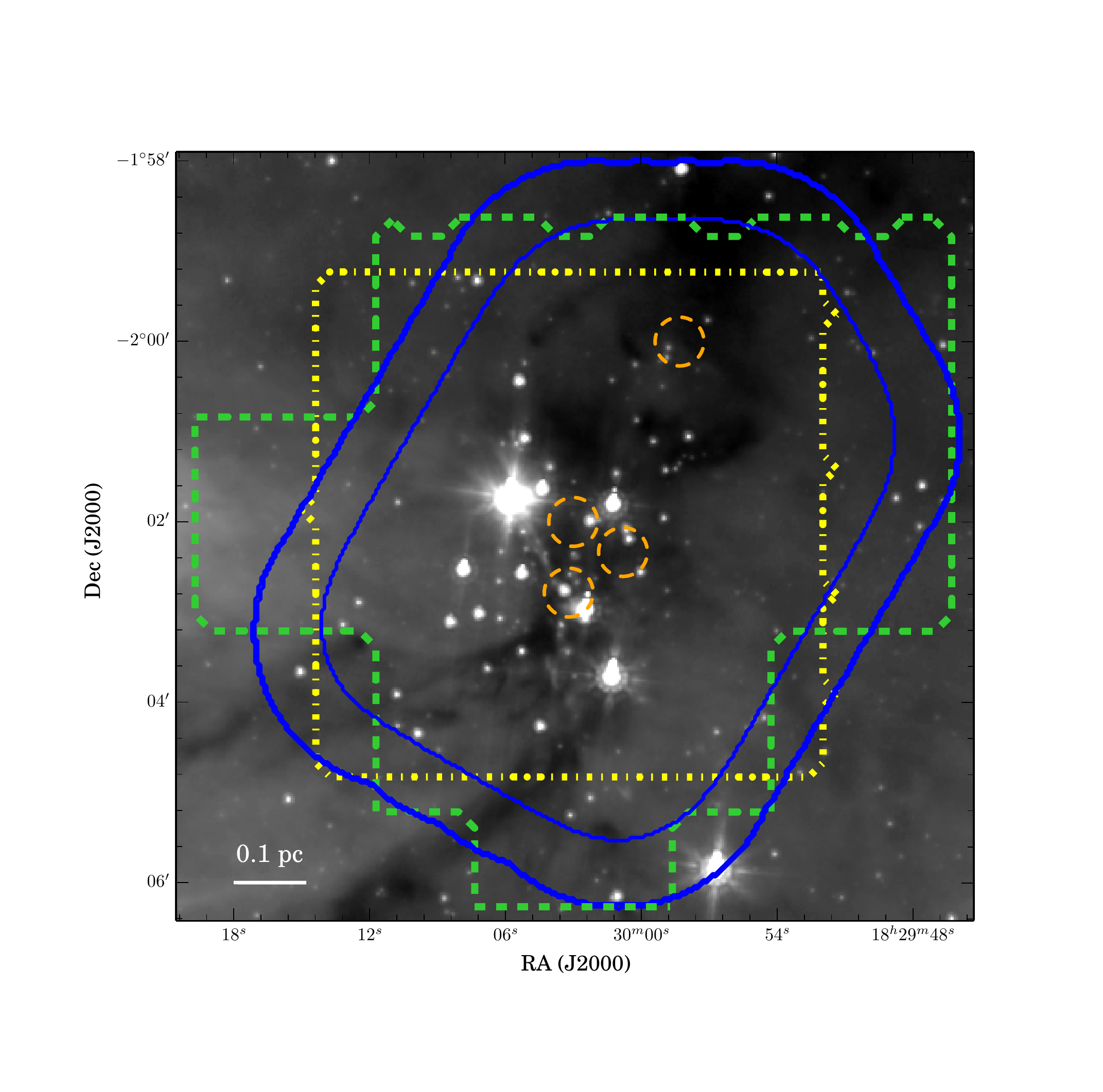} 
\caption{Coverage of observations, overlaid on map of \textit{Spitzer} 8$\mu$m emission \citep{Gut08}. Blue (solid) inner contour marks the CARMA mosaic map where Nyquist sampling and best sensitivity were achieved, and blue (solid) outer contour marks degraded sensitivity at the edge of the map.  The inner contour of this map comprises the majority of molecular outflow emission.  Green (dashed) contour marks the IRAM 30m map coverage, which is composed of nine $2\arcmin\times2\arcmin$ and one $2\arcmin\times1\arcmin$ OTF maps. Yellow (dash-dot) contour marks the APEX \twco $J=3-2$\ map.  Orange (dashed) circles mark four CSO pointings of \thco $J=3-2$\ observations, and diameters correspond to the CSO beam. }  
\label{fig:cover}
\end{figure}

The methods and results we present here pertain to data obtained with both interferometry and single dish telescopes, including detections of two energy transitions of each molecular isotopologue.  In the following sections we describe the specific observations with the Combined Array for Research in Millimeter-wave Astronomy (CARMA), IRAM 30 m, Atacama Pathfinder EXperiment (APEX), and Caltech Submillimeter Observatory (CSO), and the process of combining interferometer and single dish data of the same isotopologue.   To put the observations into context, in Figure \ref{fig:cover} we show the map coverages that we describe here; Table \ref{tab:obs} gives a summary of the observations.

We also observed the $J=3-2$\ energy level transition of $^{13}$CO with the CSO, and $^{12}$CO with the APEX.  Observations of this higher energy transition are utilized to determine velocity-dependent opacity described in Appendix \ref{sec:opacity}, and excitation temperature according to the method described in Appendix \ref{sec:tex}.

\subsection{CARMA Observations} \label{sec:CARMA}
Interferometric observations were made with CARMA in the D-array configuration during the period 2011 April and May.  We mosaicked a region of approximately $5\arcmin\times 7\arcmin$ (oriented with position angle $-30\dg$, see Figure \ref{fig:cover}) within Serpens South.  Our mosaic was centered at R.A.$=18^h30^m00^s$, decl.$=-2\dg2\arcmin0\arcsec$ and consisted of 66 pointings in a hexagonal-packed pattern; the map achieved better-than Nyquist sampling with horizontal spacing of $30\arcsec$\ and vertical spacing of $25\farcsec8$. In total, we observed 33.8 hours, with 24.6 hours on source.  Integrations were 30 seconds per mosaic position, and 3 minutes on the phase calibrator 1743-038, with 33 pointings observed in between each phase calibrator observation.  Uranus was the primary flux calibrator.

We simultaneously observed the $J=1-0$\ transitions of $^{12}$CO (115.27 GHz) and $^{13}$CO (110.20 GHz) in the upper side band (USB).  The spectral windows of $^{12}$CO and $^{13}$CO had widths of 31 MHz ($\sim 80\ \kms$) and 8 MHz ($\sim 20\ \kms$), respectively.  With 383 channels in each CO window, the velocity resolutions were 0.2 and 0.06 \kms, respectively.  In addition, we simultaneously observed the continuum emission using eight spectral windows of 495 MHz each, resulting in a map with a total continuum bandwidth of 4 GHz that covers the same region as that of CO.  The beamsize of $\sim5\as$ corresponds to a physical resolution of 0.01 pc, or 2000 AU (at a distance of 429 pc).

Data were reduced using MIRIAD \citep{Sau95}. Anomalous amplitudes and phases were flagged, as were any observations with system temperatures higher than 500 K.  No shadowed data were used, nor were observations when the source was at elevations between 85\dg and 90\dg.   Flux and gain calibrations were applied, and the data were Fourier transformed.  

Images of the data were obtained using the task \texttt{invert}, specifying a cell size of 2\arcsec, and then deconvolved with the maximum entropy deconvolution \texttt{mosmem}.  The clean images were then restored with a Gaussian beam.  We subtracted the continuum image from the molecular line maps using \texttt{avmaths} in the image domain.  A summary of the CARMA observations is given in Table \ref{tab:obs}.  

\begin{deluxetable*}{lccccccccc}
\tabletypesize{\tiny}
\setlength{\tabcolsep}{0.02in} 
\tablecaption{Observations Summary \label{tab:obs}}
\tablehead{\colhead{Line}  & \colhead{Telescope} & \colhead{Date} & \colhead{Rest Frequency}  & \multicolumn{2}{c}{HPBW} & \colhead{Beam PA}  & \colhead{Channel Width} &\colhead{Bandwidth}& \colhead{rms Noise} \\ 
\cline{5-6} 
\colhead{} & \colhead{} & \colhead{} &\colhead{(GHz)}  &  \colhead{maj($^{\prime\prime}$)} & \colhead{min ($^{\prime\prime}$)} & \colhead{($^\circ$)} & \colhead{(km s$^{-1}$)} & \colhead{(MHz)}& \colhead{} } 
\startdata
$^{12}$CO $J=1-0$   & CARMA & 2011 Apr, May &  115.27       & 5.3 &4.1  & 7.4 & 0.2  & 31 & 0.2 Jy beam$^{-1}$ \\
$^{13}$CO $J=1-0$   & CARMA & 2011 Apr, May &  110.20       & 5.3 &4.5  & -19.5 & 0.06 & 8 & 0.3 Jy beam$^{-1}$\\
$^{12}$CO $J=1-0$   & IRAM & 2012 Apr &  115.3       & 22.5 & 22.5  & \nodata & 0.53  & 4000\tablenotemark{a} & 0.2 K \\
$^{13}$CO $J=1-0$   & IRAM & 2012 Apr &  110.2       & 23.5 & 23.5  & \nodata & 0.53  & 4000\tablenotemark{a} & 0.05 K  \\
$^{12}$CO $J=3-2$   & APEX & 2009 Oct &  345.8       & 19.2 & 19.2  & \nodata  & 0.2 & 180  & 0.2 K  \\
$^{13}$CO $J=3-2$   & CSO & 2011 Jun &  330.6      & 35 & 35  & \nodata  & 0.06 & 500 & 0.4 K  \\
2.7mm continuum & CARMA & 2011 Apr, May & 110.7 &  5.3&  4.5 &  0.54& & 3958  & 0.001 Jy beam$^{-1}$\\
\enddata
\tablenotetext{a}{Total bandwidth of the observations.  Only 1024 channels centered at $v_{LSR}=10\ \kms$\ were inspected.}
\end{deluxetable*}

\subsection{IRAM Observations} \label{sec:IRAM}

We mapped the same region with the IRAM 30 m telescope in 2012 April.  The center of the map is R.A.$=18^h30^m03^s$, decl.$=-02\dg02\arcmin$ to match the CARMA map coverage.  We mapped the region using individual $2\arcmin\times 2\arcmin$ on-the-fly (OTF) maps, with each OTF map scanned twice (in orthogonal directions).  The $2\arcmin\times 2\arcmin$ maps cover a contiguous region of 38 arcmin$^2$\ outlined in Figure \ref{fig:cover}.  We tested several off positions, with the best being located at R.A.$=18^h27^m3\fs11$, decl.$=-02\dg22\arcmin00\arcsec$ (an offset of $-2700$\arcsec\ in R.A. and $+1200$\arcsec\ in decl. from the central map position).
  
Here we present data obtained using the Eight MIxer Receiver \citep[EMIR,][]{Car12} with the E090 band and the backend FTS.  We tuned to 110.5 GHz in the upper-inner sub-band to detect \thco\ $J=1-0$, and we simultaneously observed \twco\ $J=1-0$\ in the upper-outer sub-band.  The velocity resolution was  $\Delta v=0.53$\ \kms, and bandwidth was 4000 MHz (or approximately 10,000 \kms).  The beamsizes at the frequencies of \twco\ $J=1-0$\ and \thco\ $J=1-0$\ observed with the 30m telescope are 22\farcsec5 and 23\farcsec5, respectively, corresponding to $\sim0.05$\ pc at $d=429$ pc.  

We reduced the data using the GILDAS/CLASS software package (the Grenoble Image and Line Data Analysis System).\footnotemark\  For \twco\ $J=1-0$\ we corrected for beam efficiency of $\eta=0.78$ and forward efficiency of 0.95, and for \thco\ $J=1-0$\ we corrected for beam efficiency of $\eta=0.79$ and forward efficiency of 0.95 \citep{Kra13}.  We fit a first order baseline to all spectra, and combined all OTF pointings within 12\arcsec.  The resulting rms noise level of the final reduced maps are 0.2 and 0.05 K per velocity channel for  $^{12}$CO and  $^{13}$CO maps, respectively.

Deep observations (rms$\approx$0.1 K for \twco, and 0.04 for \thco) at the off position, using the frequency-switch mode, showed that there was in fact significant ($\sim2$\ K, with a width of $\sim1\ \kms$\ for \twco) extended emission in the off position at cloud velocities.  We added the frequency-switched spectrum from the off-source position to correct the OTF maps at these velocities.  Although the emission at the off position may marginally affect the spectra in our map at cloud velocities, at velocities beyond $\pm$3 \kms\ with respect to the ambient cloud velocity (the velocities we consider for outflows) we do not detect significant emission at the off position.

\footnotetext{http://www.iram.fr/IRAMFR/GILDAS}

\subsection{APEX Observations} \label{sec:APEX}
The $J=3-2$\ transition of $^{12}$CO was observed with APEX in 2009 October.  The $\sim6\arcmin\times 6\arcmin$ region centered at R.A.$=18^h30^m03^s$, decl.$=-02\dg02\arcmin$ (see Figures \ref{fig:cover} and \ref{fig:carma_apex}) was mapped using the OTF method.  The velocity resolution of the observation was $\Delta v = 0.2\ \kms$, with a bandwidth of 180 MHz, or 160\ \kms.  The beamsize at 345.8 GHz was $19\farcsec2$, and the data cube has a pixel size of 9\as.  

For the purpose of obtaining the opacity for each velocity channel (see Appendix \ref{sec:opacity}), the cube was smoothed to have a beamsize of $34\farcsec8$, and regridded to a pixel size of 11\as\ to match the CSO observations (see below).  The rms of the smoothed data cube is 0.16 K per 0.2 \kms\ channel.

These data were also used to determine excitation temperatures (see Appendix \ref{sec:tex}) in conjunction with the IRAM $^{12}$CO $J=1-0$\ data, and for this purpose the APEX data were smoothed to a beamsize of $22\farcsec5$\ and regridded to a pixel size of 11\as\ and velocity resolution of $\Delta v=0.53$\ \kms. The resultant rms following this smooth and regrid was 0.13 K per 0.53 \kms\ channel.

\subsection{CSO Observations} \label{sec:CSO}
The $J=3-2$\ transition of $^{13}$CO was observed with CSO in 2011 June.  Four positions were observed, with the central pointing being RA=$18^h30^m03^s$, Dec=$-02\dg02\arcmin$ (see Figure \ref{fig:cover}).  The additional three pointings had offsets of (+3\as,-47\as), (-70$\farcsec$5,+120\as), (-33\as,-20\as) with respect to the central pointing, selected based on various features seen in \twco\ emission.  The beamsize at 330.6 GHz is 35\as.  Velocity resolution is $\Delta v=0.06$\ \kms, and the bandwidth was 500 MHz, or 450 \kms.  

For the purpose of obtaining the opacity for each velocity channel, the cube was regridded to match the velocity resolution of the APEX observations, $\Delta v=0.2$\ \kms.  The resulting rms is 0.3 K.

\begin{figure*}[!ht]
\includegraphics[width=\linewidth]{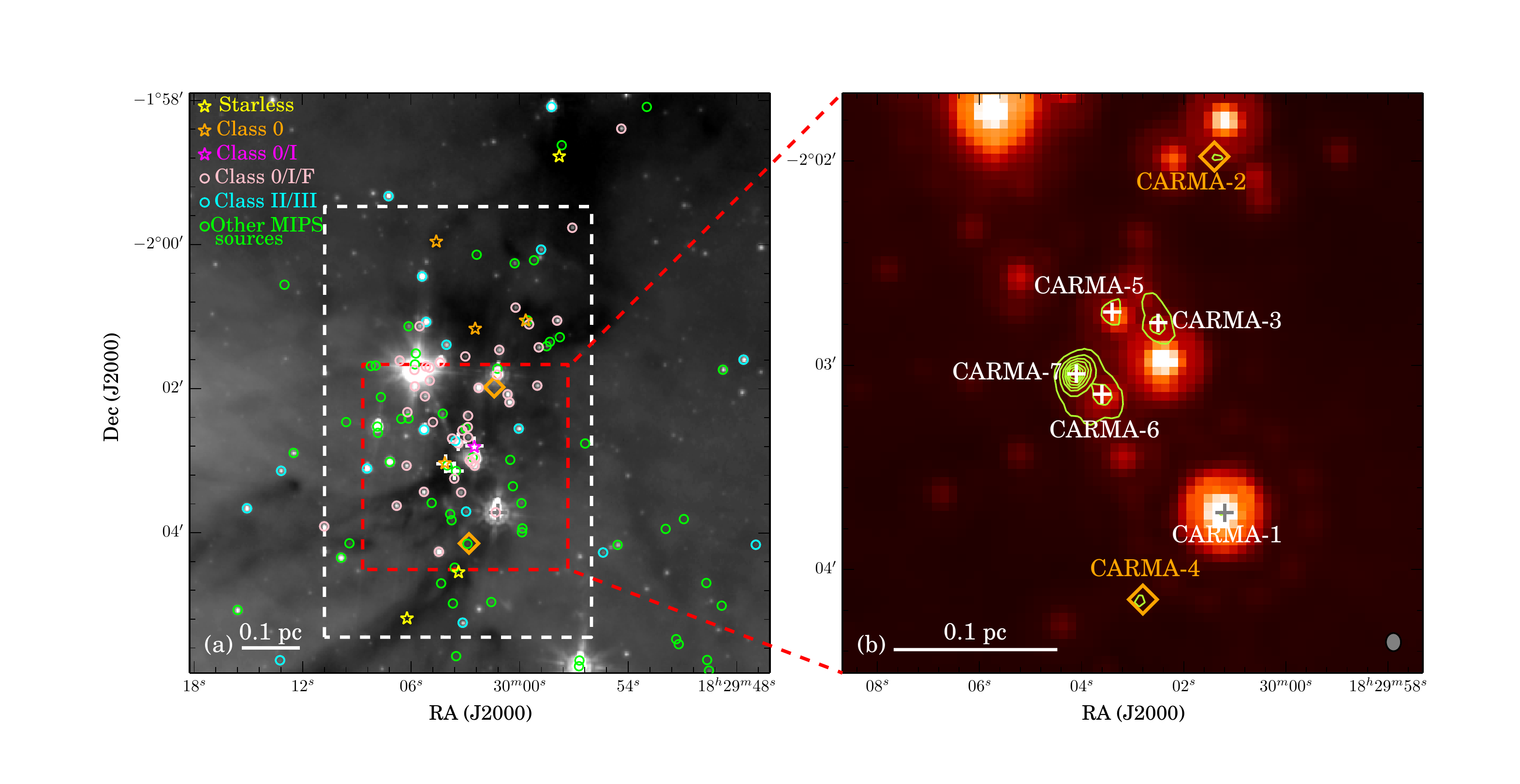}
\caption{(a) Serpens South region as observed with \textit{Spitzer} IRAC 8$\mu$m \citep{Gut08}, overlaid with sources from the literature according to their classifications by \citet[shown with star symbols]{Mau11} and in the \textit{Spitzer} Gould Belt survey \citep[][shown with circles]{Eva07,Gut08}.  The white dashed box marks the region where the majority of CO outflows are found, shown in Figure \ref{fig:outflows}.  The red dashed box shows the region where we detect continuum sources, shown in more detail in the right-hand panel.  The 2.7 mm continuum sources (see Table \ref{tab:contsources}) are given by crosses and diamonds, such that crosses indicate previously identified YSO candidates \citep{Eva07, Mau11}, and orange diamonds mark CARMA-2 and CARMA-4 that previously were not identified as YSO candidates.  (b) 2.7 mm continuum (contours) observed with CARMA, overlaid on the \textit{Spitzer} 24$\mu$m map \citep{Gut08}. Noise rms of the continuum map is 1 mJy/beam, and the beam is shown in the lower right corner.  Contours begin at a level of $4\sigma$ and increment by $8\sigma$.  Overlaid are the positions of 2.7 mm continuum sources, as described for the left panel.}  
\label{fig:continuum}
\end{figure*}

\subsection{Combining Interferometer and Single Dish Data} \label{sec:combine}

Combining interferometer (CARMA) and single dish (IRAM 30 m) \twco\ and \thco\ $J=1-0$\ data recovers extended emission (up to the combined map size of about 0.8 pc) that was filtered out by the interferometry, and it is imperative for accurately determining mass, momentum and energy of the outflows, presented in Section \ref{sec:mass}.  The combination follows the same method as described in \citet{Plu13}, corresponding to the nonlinear joint deconvolution method of \citet{Sta02}.  This method is particularly useful for recovering flux in mosaics that cover regions with clumpy structure.   The combination was done with MIRIAD.  First we regridded the interferometer map to the velocity resolution of the single dish map, and then regridded the pixels of the single dish map using the interferometer map as the template.  

The joint deconvolution was done with the MIRIAD task \texttt{mosmem} using the dirty interferometer map and single dish (regridded) map as inputs, along with dirty beams for each.  The beam sizes were $5.3\as\times4.1\as$ for $^{12}$CO and $5.3\as\times4.5\as$ for $^{13}$CO.  We converted the single dish maps to have units of Jy beam$^{-1}$, in order to match the units of the interferometer maps.  Several other parameters were necessary inputs in MIRIAD so that the convolution process converged within a reasonable number of iterations.  One such input, ``rmsfac'', specifies the ratios between the true noise rms (output by MIRIAD using \texttt{imstat} in several emission-free regions) and the theoretical rms (determined by \texttt{mossen}) for both the interferometer and single dish maps.  These factors are necessary in order for \texttt{mosmem} to reduce the residuals in the combined map (i.e., the difference between the dirty image and the model modified by the point-spread function) to have the same rms as the theoretical rms multiplied by the ``rmsfac''.  We used ``rmsfac'' of 1.6 and 1.5 for the interferometer and single dish maps, respectively.  

Another parameter is the flux calibration factor, or the number by which the single dish data are multiplied in order to convert to the same scale as the interferometer data.  We determined this factor using the task \texttt{immerge}, choosing uv-data in the range $12-25$\ m (which is the range of ``baselines'' that the interferometer and single dish data have in common), and specifying channels where emission structure is comparable in the interferometer and singe dish maps.  We used a factor of 1.1 for the $^{12}$CO deconvolution, and 1.3 for the $^{13}$CO deconvolution.   Using these parameters, the deconvolution converged within 100 iterations for each channel. The final ``CLEAN'' maps were generated with the task \texttt{restor}.  The resulting mean rms sensitivity of the combined CARMA and IRAM maps of $^{12}$CO  and $^{13}$CO were 0.16 Jy/beam/channel and 0.11 Jy/beam/channel, respectively, with 0.5 \kms\ channel width and beamsize $5.3\arcsec\times4.6\arcsec$\ (corresponding to $\sim0.01$ pc at $d=429$ pc).  The zeroth-order moment map of the combined CARMA+IRAM \twco\ data is shown in Figure \ref{fig:carma_apex} (without including the ambient velocity channels, and where the blue and redshifted wing emission are shown in contours), and channel maps of \twco\ and \thco\ are shown in Figures \ref{fig:channel2}-\ref{fig:channel2b}, respectively.

\begin{deluxetable*}{lccccccccccl}
\tabletypesize{\tiny}
\setlength{\tabcolsep}{0.01in} 
\tablecaption{Continuum sources \label{tab:contsources}}
\tablehead{ \colhead{Source} & \multicolumn{2}{c}{Position}&& \multicolumn{2}{c}{Size}&\colhead{PA}  &\colhead{Peak Intensity}&\colhead{Total Flux} & \colhead{Mass} & \colhead{c2d \tablenotemark{a}} & \colhead{Other}\\
\cline{2-3} \cline{5-6} 
\colhead{}&\colhead{$\alpha$(J2000)}&\colhead{$\delta$(J2000)}  && \colhead{Maj (\arcsec)}& \colhead{Min (\arcsec)}& \colhead{(\dg)}& \colhead{(mJy/beam)}&\colhead{(mJy)} &\colhead{(\msun)} & \colhead{Class.}& \colhead{Names} }
\startdata
CARMA-1 & 18:30:01.2 & -02:03:43.3 && 6.6 & 2.9 & 38.9 & 3.6$\pm$0.8 & 7.1 & 0.07 & YSOc & 4 \tablenotemark{b}, Y1 \tablenotemark{c} \\ 
CARMA-2 & 18:30:01.4 & -02:01:58.7 && 6.5 & 3.3 & -61.6 & 3.6$\pm$1.3 & 7.5 & 0.07 & \nodata& \nodata \\ 
CARMA-3 \tablenotemark{d}& 18:30:02.5 & -02:02:47.5 && 8.0 & 3.2 & 18.6 & 17.5$\pm$2.4 & 39.1 & 0.36 & \nodata& SerpS-MM16 \tablenotemark{e}  \\ 
\hline
CARMA-3 SW \tablenotemark{d} & 18:30:02.5 & -02:02:48.3 && 4.0 & 3.0 & -87.4 & 19.1$\pm$2.4 & 29.5 & 0.27 & \nodata& \nodata  \\ 
CARMA-3 NE \tablenotemark{d} & 18:30:02.6 & -02:02:41.4 && 4.0 & 3.1 & 80.1 & 6.4$\pm$2.1 & 10.0 & 0.09 & \nodata& \nodata  \\ 
\hline
CARMA-4& 18:30:02.8 & -02:04:08.8 && \nodata \tablenotemark{f} & \nodata &\nodata & 6.2$\pm$1.7 & 7.1 & 0.07 \tablenotemark{g} & Galc\_red & \nodata  \\ 
CARMA-5& 18:30:03.4 & -02:02:44.3 && 5.6 & 3.6 & -14.8 & 7.0$\pm$1.1 & 13.1 & 0.12& YSOc\_red  & 1 \tablenotemark{b}, P2 \tablenotemark{c} \\ 
CARMA-6 & 18:30:03.6 & -02:03:08.5 && 7.0 & 4.6 & 36.7 & 19.4$\pm$1.5 & 46.5 & 0.43 & red & SerpS-MM18 \tablenotemark{h},  \\ 
 &  & &&  &  & & &  &  & & 2 \tablenotemark{h}, P3 \tablenotemark{c}  \\ 
CARMA-7 & 18:30:04.1 & -02:03:02.5 && 6.0 & 4.1 & -32.9 & 63.6$\pm$4.4 & 131.4 & 1.21 & red & SerpS-MM18 \tablenotemark{h}, \\
 &  & &&  &  & & &  &  & &  2 \tablenotemark{h} 
\enddata
\tablecomments{Position, size, position angle (PA), peak intensity, and total flux from Miriad task ``imfit'' with Gaussian fit.  Size and PA are the deconvolved fits. }
\tablenotetext{a}{\citet{Eva07} documents the c2d data pipeline and classification, and \textit{Spitzer} data are presented by \citet{Gut08}.}
\tablenotetext{b}{From Figure 11 of \citet{Nak11}, although exact coordinates are not given.}
\tablenotetext{c}{\citet{Tei12}.}
\tablenotetext{d}{CARMA-3 was fit with a single Gaussian, while CARMA-3 SW and NE correspond to the same system fit with a double Gaussian.}
\tablenotetext{e}{\citet{Mau11} classify MM16 as  Class 0/I, integrated over a multiple system.}
\tablenotetext{f}{Deconvolution with Miriad task ``imfit'' fails.}
\tablenotetext{g}{We present a mass for this object as if it were an embedded source within Serpens South, although this source was classified as a galaxy candidate according to \citep{Eva07}.}
\tablenotetext{h}{CARMA-6 and CARMA-7 correspond to two MIPS sources, and are identified as a multiple system known as SerpS-MM18 by \citet{Mau11} and source 2 by \citet{Nak11}.  }
\end{deluxetable*}

\section{Results}\label{sec:results}

\subsection{Continuum Sources} \label{sec:continuumsources}
Using our CARMA continuum observations, we detected seven distinct continuum sources with flux greater than $4\sigma$, and we labeled them in numerical order with increasing R.A.  The sources are shown in Figure \ref{fig:continuum} and given in Table \ref{tab:contsources}.  These sources were fit as 2D Gaussians using the MIRIAD task \texttt{imfit}.  Using the integrated flux measured with \texttt{imfit}, we calculated masses of the continuum sources according to the following relation \citep[see][]{Sch10}:
\begin{equation}
M=\frac{d^2 S_\nu}{B_\nu(T_D) \kappa_\nu},
\end{equation}
where $d$ is distance to the source,  $S_\nu$ is the 2.7 mm continuum flux, and $B_\nu(T_D)$ is the Planck function.  We assumed a dust temperature of $T_D=30$ K \citep{Che13}, and a dust opacity of $\kappa_\nu=0.1(\nu/1200\textrm{ GHz})^\beta$\ cm$^2$g$^{-1}$, with $\beta=1$, which corresponds to $\kappa_\nu=0.0092$\ cm$^2$g$^{-1}$ at $\lambda=2.7$\ mm \citep{Loo00}.  Using the dust opacities extrapolated from \citet[][Column 6 of Table 1]{Oss94}, assuming a gas-to-dust ratio of 100 and $\beta=1$, then $\kappa^{OH}_\nu(\beta=1)=0.0043$\ cm$^2$\ g$^{-1}$; extrapolating instead using $\beta=2.0$, then $\kappa^{OH}_\nu(\beta=2.0)=0.0021$\ cm$^2$\ g$^{-1}$.  In these cases, continuum masses would be about 2-4 times less than what we report, respectively.  We note these specifically to remind that without further detailed modeling, our simple mass calculation is uncertain by a factor of a few depending on choices of $\kappa$\ and its spectral index $\beta$.  \citet{Sch14} found a median emissivity spectral index of $\beta=0.9\pm0.3$, and so our assumption of $\beta=1$ is within their uncertainty.

Given that our continuum map has an rms of 1 mJy, we are able to detect envelopes that have masses greater than 0.04 \msun\ (at better than a $4\sigma$ level)\todo{Value}.  Of the continuum sources in our map, the sources CARMA-3 and CARMA-6/7 were previously detected in the MAMBO 1.2 mm dust continuum map of \citet{Mau11}, and they also appear to coincide with the central clump detected in the ASTE/AzTEC 1.1 mm continuum emission map \citep[see e.g.,][]{Kir13,Tan13}. \citet{Mau11} classified these sources (which they call SerpS-MM16 and SerpS-MM18) as Class 0/I and Class 0, respectively, based on the envelope mass versus bolometric luminosity ($M_{env}-L_{bol}$) evolutionary diagram that they present.  Further, \citet{Mau11} noted that these particular sources are each multiples, although they report quantities that are integrated over each system.  

The extended morphologies shown in our 2.7 mm continuum map (see Figure \ref{fig:continuum}) reinforce the notion that the sources CARMA-3 and CARMA-6/7 are each multiples.  The source CARMA-3 is seen as an elongated structure with a peak of emission in the southwest and extended emission to the northeast.  According to a double-Gaussian fit, the centers of two components in this system are separated by $\sim7\arcsec$, or $\sim3000$\ AU not taking into account inclination effects. The total integrated flux of the sources is 39 mJy, with the southwest (CARMA-3 SW) and northeast (CARMA-3 NE) components representing 75 and 25\% of the flux, respectively.  These components likely share a circumbinary envelope, and since there is no absolute minimum in emission marking a clear delineation between the two components, in Table \ref{tab:contsources} we also present a single-Gaussian fit and the corresponding total mass for this system.  A better understanding of the individual components merits further modeling.

For the source CARMA-6/7, although the multiple components show some overlapping extended emission, we see two distinct peaks of emission, with the northeastern peak (CARMA-7) being the strongest continuum emission in our map.  The peaks of emission are separated by $\sim 16\arcsec$\ ($\sim6500$ AU) on the plane of the sky, but dust emission between the sources may pertain to a shared circum-binary envelope. Since we detect two distinct peaks, we fit two Gaussians with \texttt{imfit}, and we determine the masses of the components CARMA-6 and CARMA-7 to be 0.43 and 1.21 \msun, respectively. \todo{Value} 

Two distinct $\textit{Spitzer}$ MIPS 24 $\mu$m sources are coincident with the two peaks of mm continuum emission associated with CARMA-6/7, and the MIPS sources were classified as ``red'' following \citet{Eva07}.  The label ``red'' indicates that the flux density at 24 $\mu$m is at least three times the flux density of the nearest available IRAC band detection in \citet{Gut08}.  We suggest that since the MIPS 24 $\mu$m sources coincide with the continuum sources, they are likely deeply embedded YSO candidates, even though these sources lack detection in one of the IRAC bands and therefore are not classified as YSO candidates according to the criteria of \citet{Eva07}.  CARMA-3 lies about 10\as\ north of several \textit{Spitzer} sources, and at least one of those is particularly bright in 4 $\mu$m and 8 $\mu$m emission, as can be seen in Figure \ref{fig:continuum}.  Upon close inspection of the location of this continuum source, we find weak $4-24$\ $\mu$m emission at the position of CARMA-3, but it is not clear whether the $4-24$\ $\mu$m emission is associated with a stronger source to the south.  This may have detracted from detecting a source at the location of CARMA-3 according to the criteria of \citet{Eva07}, as emission from a stronger source could have dominated an underlying weaker source nearby.  Additionally, a source seen in mm continuum emission but not 24 $\mu$m emission may signify that this source is still IR-dark and therefore at a very early stage of evolution.

Three continuum sources in our map have not been previously published as mm continuum sources, but do coincide with MIPS 24 $\mu$m sources.  These sources we name CARMA-1, CARMA-4 and CARMA-5 (see Table \ref{tab:contsources}).  CARMA-1 and CARMA-5 were classified as YSOc and YSOc\_red \citep{Eva07}, respectively, indicating YSO candidates (``red'' is described above).  CARMA-5 is located only $\sim15\arcsec$ (0.03 pc) from CARMA-3, making it difficult to resolve with previous single dish observations.  Among the continuum sources in this region, CARMA-1 has the largest corresponding MIPS 24 $\mu$m flux.  The combination of large 24 $\mu$m flux and small mass (based on 2.7 mm continuum flux) suggest that this source is among the more evolved mm-sources in the map.

Several of the continuum sources that we have discussed here have been referred to by various names in previous literature, and we refer the reader to the right column of Table \ref{tab:contsources} for these labels.  In addition, \citet{Tei12} indicate an IR source in the region to the northeast which they call P4 and they claim drives an outflow, but we detect no continuum counterpart for this source.  To our knowledge, the source CARMA-2 has not been identified in previous observations.  This source only has two pixels with continuum flux greater than $4\sigma$, but these are surrounded by pixels with flux of $2-3\sigma$\ (hence, only marginally detected above the rms noise).  If this source is indeed a YSO in Serpens South, we calculate a mass of 0.07 \msun.\todo{Value}

Finally, we also detect unresolved continuum emission associated with the source CARMA-4, for which we cannot confirm whether it is a YSO candidate or a background galaxy detection.  First of all, it was classified as a galaxy candidate (``Galc\_red'') according to \citet{Eva07}.  This classification was due to its weak MIPS 24 $\mu$m detection of $9.91\pm1.05$ mJy, which is about a factor of at least 10 weaker than the other MIPS sources previously described in this region.  However, the coincident continuum emission may be evidence that this is a YSO candidate.  If we assume that this source is indeed a YSO in Serpens South, the continuum emission implies an inner envelope/disk mass of 0.07 \msun\ (see Table \ref{tab:contsources}). \todo{Value}

In order to compare with sources detected (or not detected) in previous mm-continuum observations of this region by \citet{Mau11}, we scaled the continuum emission fluxes detected at 2.7 mm (110 GHz) to the expected flux at 1.2 mm (240 GHz).  In that case, we expect thermal emission flux to increase by a factor of $9 - 20$\ for $\beta = 1 - 2$, respectively.  For CARMA-3 and CARMA-6/7 -- the sources that we detect in common with \citet{Mau11} -- we measure integrated fluxes at 2.7 mm that, when scaled to 1.2 mm assuming $\beta = 1$, correspond to 80\% and 90\%, respectively, of the fluxes measured by \citet{Mau11}.  This agreement within uncertainties of the 2.7 mm and 1.2 mm continuum fluxes, made with interferometer and single dish telescope observations, respectively, suggests: (1) the majority of emission for these sources arises from thermal dust emission on the Rayleigh-Jeans tail \citep[also described in the Orion Molecular Cloud by][]{Sch14}, and (2) the sources are compact sources without significant extended emission that would be filtered out by the interferometer observations.  Conversely, assuming $\beta = 2$ in the scaling from 2.7 mm to 1.2 mm results in expected 1.2 mm integrated fluxes that are greater by about a factor of two than those reported by \citet{Mau11}.  Without a more detailed study to determine the actual spectral emissivity of these sources, we cannot rule out the possibility that up to half of the 2.7 mm flux that we detect is contributed from non-thermal emission, such as free-free emission.

We also investigate the remaining sources that were not detected in 1.2 mm observations by \citet{Mau11}.  Specifically, we detected the sources CARMA-1, CARMA-2, CARMA-4, and CARMA-5 with a peak intensity $\leq7\sigma$, and we scaled the 2.7 mm fluxes to expected 1.2 mm fluxes for comparison.  Further, accounting for beam dilution in an 11\arcsec\ beam, the corresponding flux detected by \citet{Mau11} would be reduced by at least a factor of about 5 (assuming that an unresolved source fully fills our $5\arcsec$\ beam, which we consider an upper limit for what are likely more compact sources).  Therefore, we find that a $1\sigma$ detection in our map corresponds to $\sim 2-5$\ mJy/11\arcsec-beam at 1.2 mm, whereas \citet{Mau11} report a $1\sigma$\ rms noise of $15$\ mJy/11\arcsec-beam.  Therefore, our 2.7 mm continuum source detections that are not in common with those of \citet{Mau11} are likely due to differences in sensitivity and beamsize of the observations; even a $7\sigma$ detection in our 2.7 mm map would correspond to $\sim1\sigma$ detection in a 1.2 mm map with an 11\arcsec\ beam.   

\begin{figure*}[!ht]
\includegraphics[width=\linewidth]{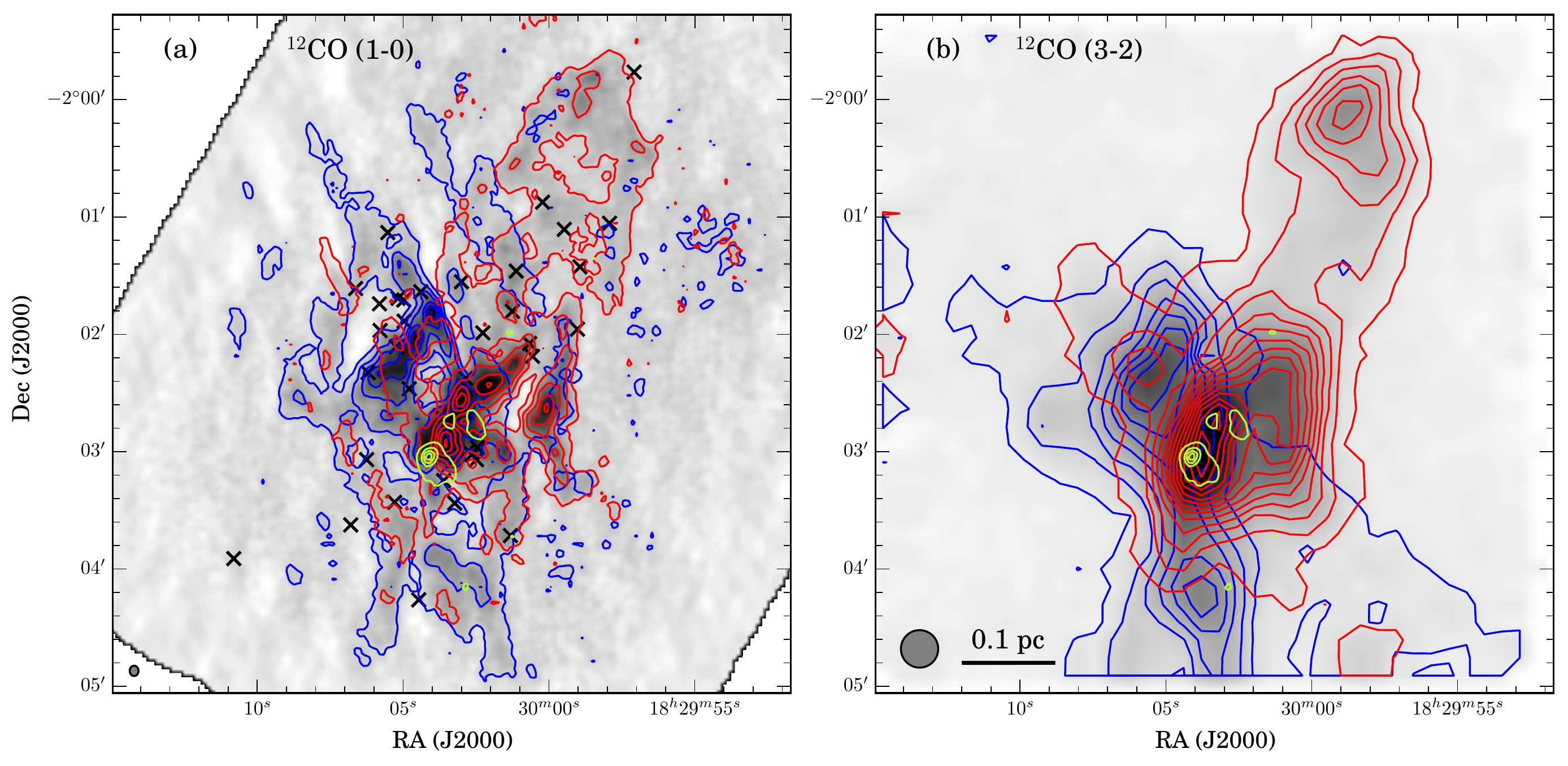} 
\caption{Integrated intensity maps of the blue and redshifted outflow emission in Serpens South traced by (a) \twco\ $J=1-0$\ emission observed with CARMA+IRAM, and (b) \twco\ $J=3-2$\ emission observed with APEX, with beam sizes for each given in lower left corners.  The scale bar applies for both panels.  Grayscale shows integrated intensity with $3<|v_{out}|<31$\ \kms, and blue and red contours show the blue and redshifted moment zero maps within this velocity range. Contours begin at $5\sigma$ for each respective map, and increase in intervals of $10\sigma$.  Green contours show 2.7 mm continuum observed with CARMA, beginning with $4\sigma$ and increasing by $8\sigma$.  Crosses mark the 45 protostars in the region from the \textit{Spitzer} Gould Belt survey \citep{Eva07,Gut08}, and these are also shown in Figure \ref{fig:continuum} according to their classifications. }  
\label{fig:carma_apex}
\end{figure*}

In the following section, we refer to several of these continuum sources as potential outflow-driving sources, and in Figure \ref{fig:carma_apex} we show continuum emission along with 0th moment CO maps.  In that figure (also Figure \ref{fig:continuum} and \ref{fig:outflowsz}), we overlay the locations of protostars in the region identified in the \textit{Spitzer} Gould Belt survey \citep{Eva07,Gut08}.  We note that while we focus on the continuum sources in the context of our discussion of possible outflow-driving sources, there are many ($\sim45$) densely packed protostars in the mapped region.  Specifically, within a radius of 30\as\ from a point centered between the sources CARMA-3, CARMA-5, CARMA-6, and CARMA-7 (the location that appears to be the approximate apex of the majority of outflow emission) reside 10 protostars identified with \textit{Spitzer}, and 18 protostars reside within a radius of 1\arcmin. Whether some \textit{Spitzer} sources correspond to weak continuum sources below our detection limit, or do not correspond to continuum sources at all, they nonetheless may contribute to the outflow activity described in \S \ref{sec:outflows}.  In a future study, we plan to present observations with higher resolution and sensitivity in the central region that will likely give us a better grasp of specific outflow-driving sources, and in that case we plan to discuss the \textit{Spitzer} and continuum sources further. 

\begin{figure*}[!ht]
\includegraphics[width=\linewidth]{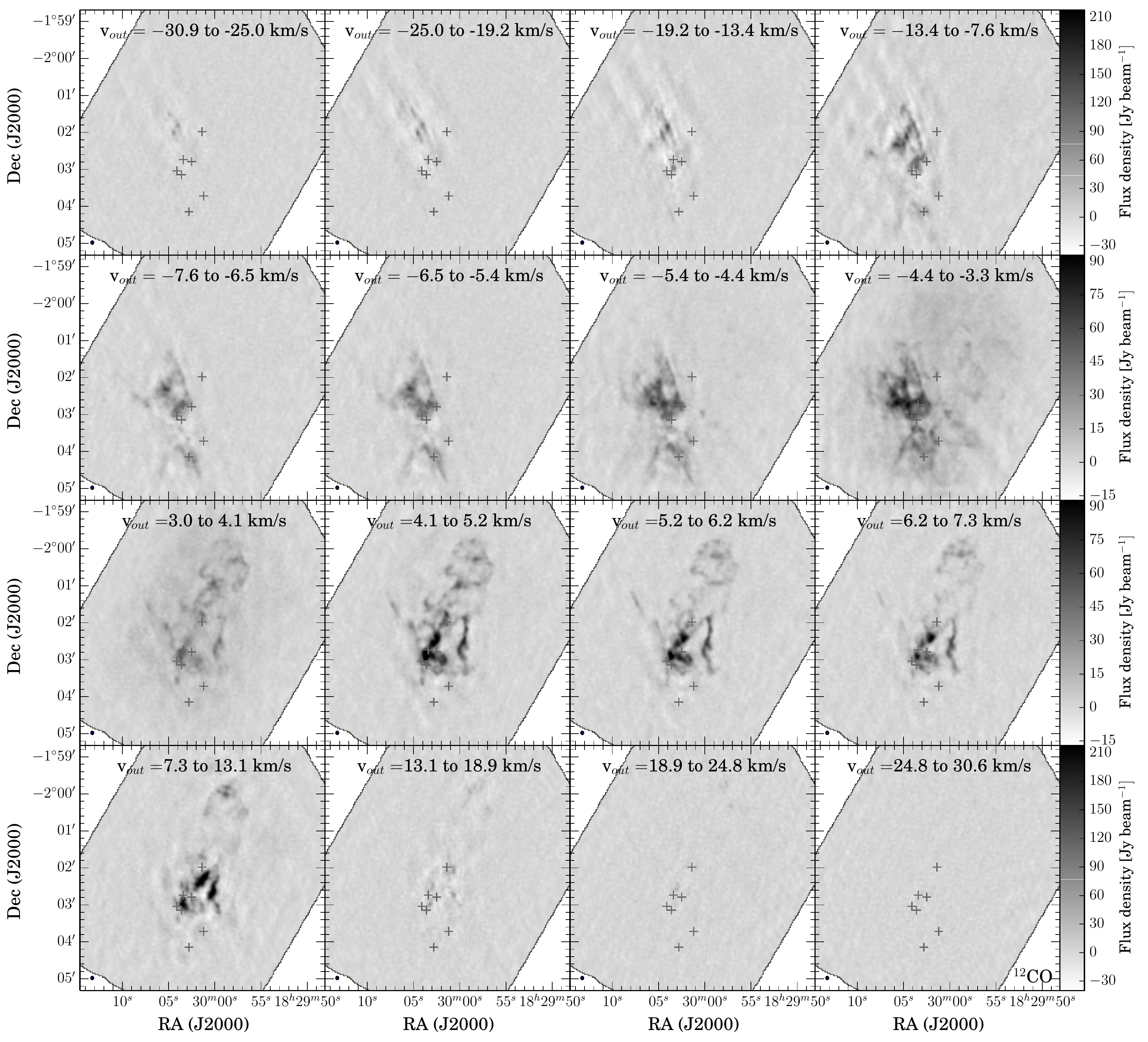}
\caption{$^{12}$CO $J=1-0$\ channel maps using combined CARMA and IRAM data, with positions of 2.7 mm continuum sources observed with CARMA marked with plus signs.  Velocity channels are grouped such that panels in top and bottom rows have velocity ranges of $\Delta v=5.8\ \kms$\ (except for upper left and lower right, which have $\Delta v=5.3\ \kms$), and panels in middle two rows have velocity ranges of $\Delta v=1.1\ \kms$.  Color scale bars are given for each corresponding row.  Specific outflow velocities are indicated in respective panels, and velocity channels within $\pm 3$\ \kms\ of the cloud velocity are omitted.  }  
\label{fig:channel2}
\end{figure*}

\begin{figure*}[!ht]
\includegraphics[width=\linewidth]{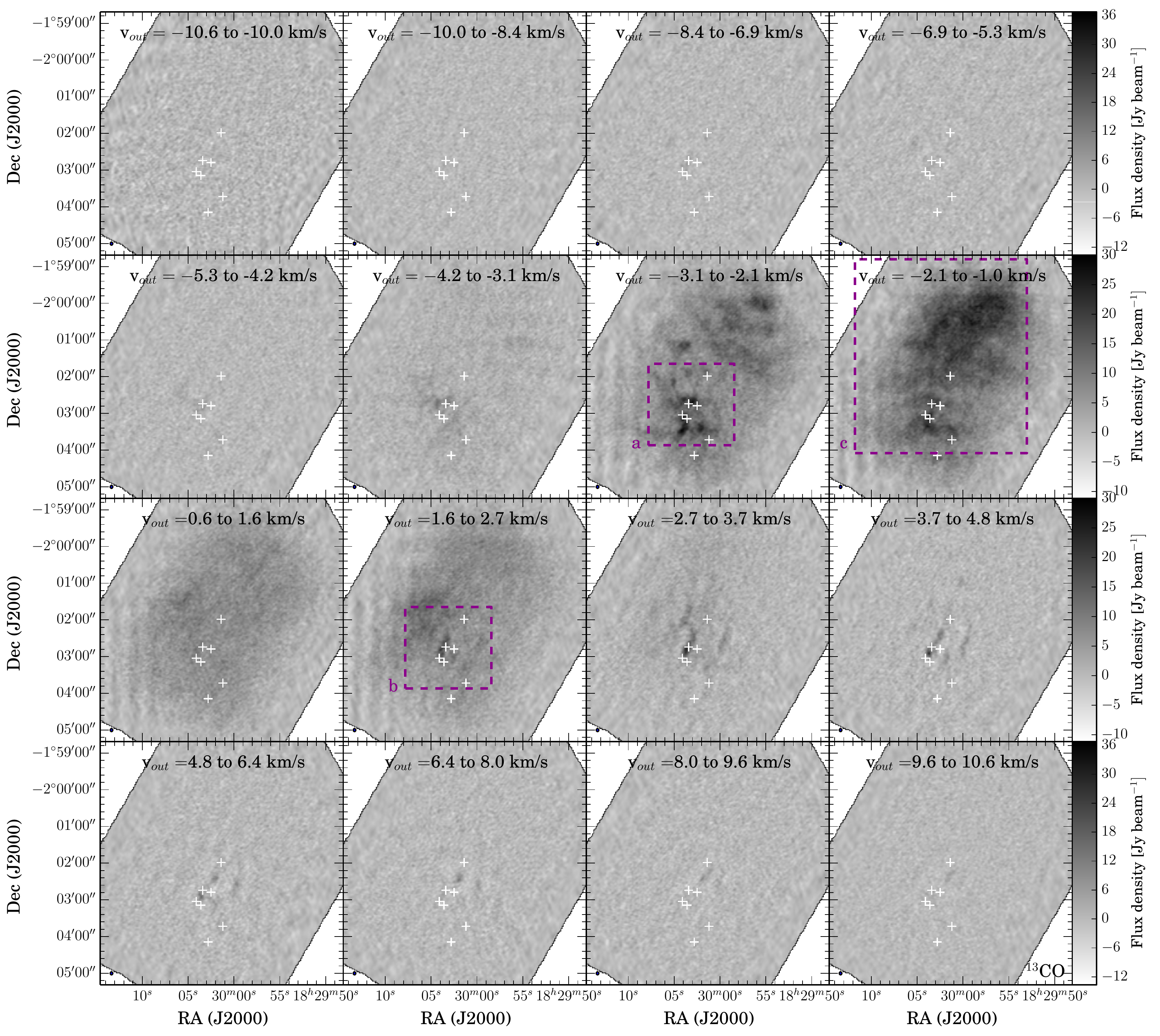}
\caption{$^{13}$CO $J=1-0$\ channel maps using combined CARMA and IRAM data (with plus signs as in Figure \ref{fig:channel2}).  Velocity channels are grouped such that panels in top and bottom rows have velocity ranges of $\Delta v=1.6\ \kms$\ (except for upper left and lower right), and panels in middle two rows have velocity ranges of $\Delta v=1.1\ \kms$.  Color scale bars are given for each corresponding row.  Specific outflow velocities are indicated in respective panels, and velocity channels within $\pm 0.6$\ \kms\ of the cloud velocity are omitted.  Additionally, dashed boxes indicate the regions shown in more detail in Figure \ref{fig:12co_13co}. }  
\label{fig:channel2b}
\end{figure*}

\subsection{CO Outflow Morphologies} \label{sec:outflows}

Our suite of high spatial-resolution mm-wavelength $^{12}$CO and continuum maps, along with archival \textit{Spitzer} IRAC images, allow us to scrutinize the outflow morphologies in the most dense regions, assess the impact of these outflows on their surrounding environment, suggest probable driving sources where possible, and compare with previous identifications from the literature.   Along with the dense concentration of YSOs in the central cluster previously mentioned, the resolution of our observations and the breadth of the outflow emission make it a challenge to identify individual driving sources.  In this section we describe the CO outflow morphologies that we observed at particular velocity ranges, and we present the quantitative characteristics of these outflows in Section \ref{sec:mass}.

We accompany the following discussion with references to several CO maps in Figures \ref{fig:carma_apex}-\ref{fig:outflow_cand1}, beginning with Figure \ref{fig:carma_apex} showing 0th moment maps of \twco\ $J=1-0$\ and $J=3-2$\ emission.  In Figure \ref{fig:channel2}, we show \twco\ $J=1-0$\ velocity-binned maps, summing channels in variable-sized bins so that near the cloud velocity we sum over two consecutive channels ($\Delta v=1.1\ \kms$), and for high outflow velocities we sum over ten channels ($\Delta v=5.8\ \kms$).  Additionally, \thco\ emission is shown in a similar manner in Figure \ref{fig:channel2b}, while featuring especially the lower-velocity channels where \thco\ emission is detected.  Figure \ref{fig:12co_13co} shows \thco\ overlaid with \twco\ contours for certain velocity ranges. Figures \ref{fig:outflows}-\ref{fig:outflowsz} are moment maps of low-, mid-, and high-velocity outflow emission probed by \twco\ $J=1-0$\ CARMA+IRAM observations.  Velocity ranges shown in those maps are asymmetric about the cloud velocity in order to reveal particular features that were identified by close inspection of the data cube, and those features are discussed specifically in the text in the following sections.  An outflow candidate is shown in a similar way (but at lower outflow velocities) in Figure \ref{fig:outflow_cand1}, and discussed in \S \ref{sec:carma1out}.

Throughout, we use the notation of $v_{LSR}$ to indicate channel velocity (relative to local standard of rest), $v_{cloud}$ for cloud velocity, and $v_{out}=v_{LSR}-v_{cloud}$ for outflow velocity (along the line of sight).  We adopt a systemic cloud velocity of $v_{cloud}=\vcloud$ pertaining to the region we mapped.  Previous works \citep[e.g.,][]{Nak11,Kir13} adopted a cloud-wide velocity centroid of $v_{cloud}=7.5\ \kms$, but Figure 4 of \citet{Kir13} shows that in the region we mapped (centered at R.A.$=18^h30^m$, decl.$=-2\dg2\arcmin$), there is a predominance of a larger centroid velocity of approximately $v_{cloud}=8\ \kms$.

\subsubsection{Blueshifted outflows} \label{sec:blueout}

\begin{figure*}[!ht]
\includegraphics[width=\linewidth]{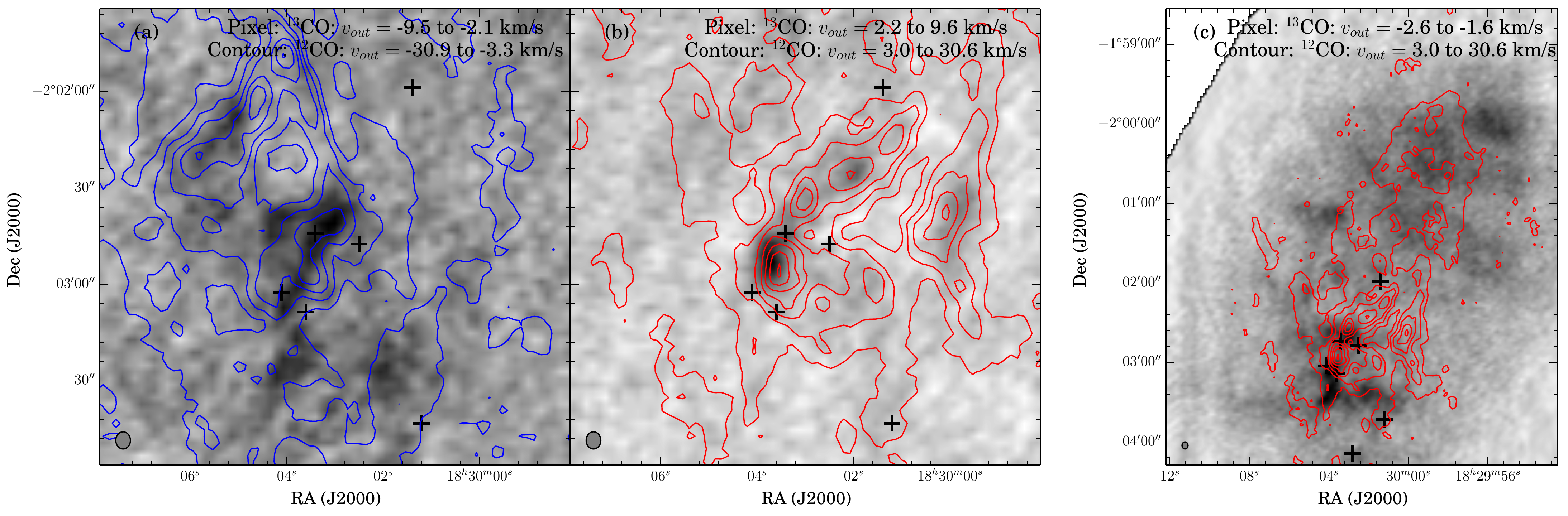}
\caption{\twco\ and \thco\ emission in several regions of the maps, showing where the low-velocity entrained gas traced by \thco\ coincides with molecular outflows traced by \twco.  Panels (a) and (b) show blue and redshifted \twco\ emission (contours) on pixel maps of \thco\ at lower velocities of $|v_{out}|<10$\ \kms.  Panel (c) shows redshifted \twco\ (contours) overlaid on lower-velocity blueshifted \thco. The contours are the same as in Figure \ref{fig:carma_apex}(a), and regions correspond to dashed boxes shown in Figure \ref{fig:channel2b}. }  
\label{fig:12co_13co}
\end{figure*}

\begin{figure*}[!ht]
\includegraphics[width=0.9\linewidth]{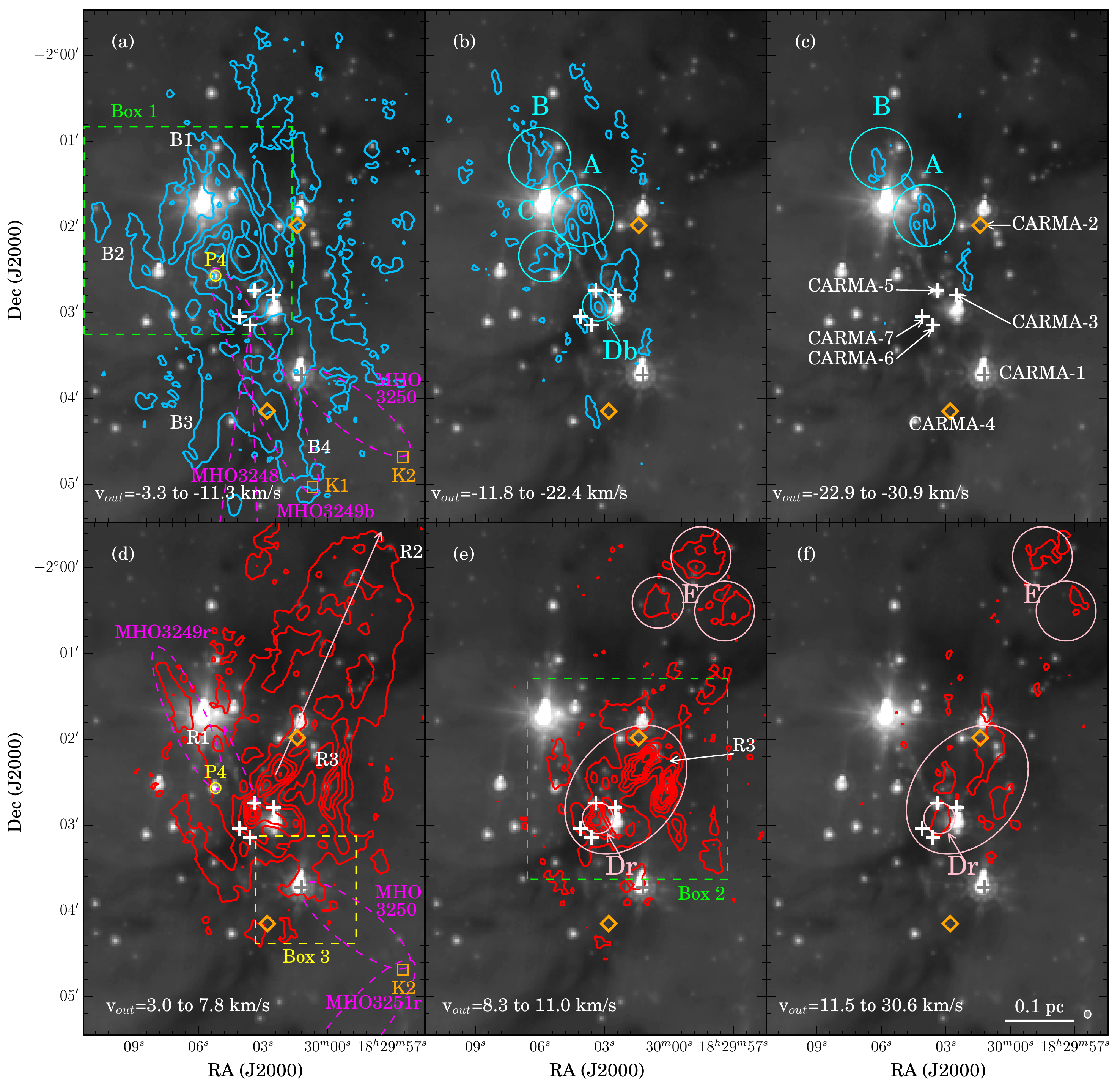}
\caption{$^{12}$CO outflow emission, showing low-velocity (left), mid-velocity (middle) and high-velocity (right) outflowing gas features.  Blue and red contours represent blue and redshifted gas, and velocity intervals are indicated for each panel, where outflow velocity, v$_{out}$, is relative to the central cloud velocity of \vcloud.  Contours begin at 5$\sigma$ for each 0th moment map, and increment by 15$\sigma$.  The beamsize and scale bar are given in the lower right corner.  Background is \textit{Spitzer} 8$\mu$m emission \citep{Gut08}.  Boxes 1, 2, and 3 refer to the regions shown in Figures \ref{fig:outflowsz}(a-c), \ref{fig:outflowsz}(d-f), and \ref{fig:outflow_cand1}, respectively. Ellipses and the arrow with solid cyan and pink lines indicate outflow features mentioned in the text, and plus signs and diamonds mark the positions of continuum sources detected by our CARMA observations, labeled in panel (c). K1 and K2 are IR H-H objects identified by \citet{Nak11}, and correspond to the southwestern tips of the molecular hydrogen objects MHO3249b and MHO3250 identified by \citet{Tei12}; P4 is an IR source.  MHO features from \citet[][see their Figure 1]{Tei12} are indicated with magenta dashed lines and ellipses. }  
\label{fig:outflows}
\end{figure*}

\begin{figure*}[!ht]
\includegraphics[height=0.44\textheight]{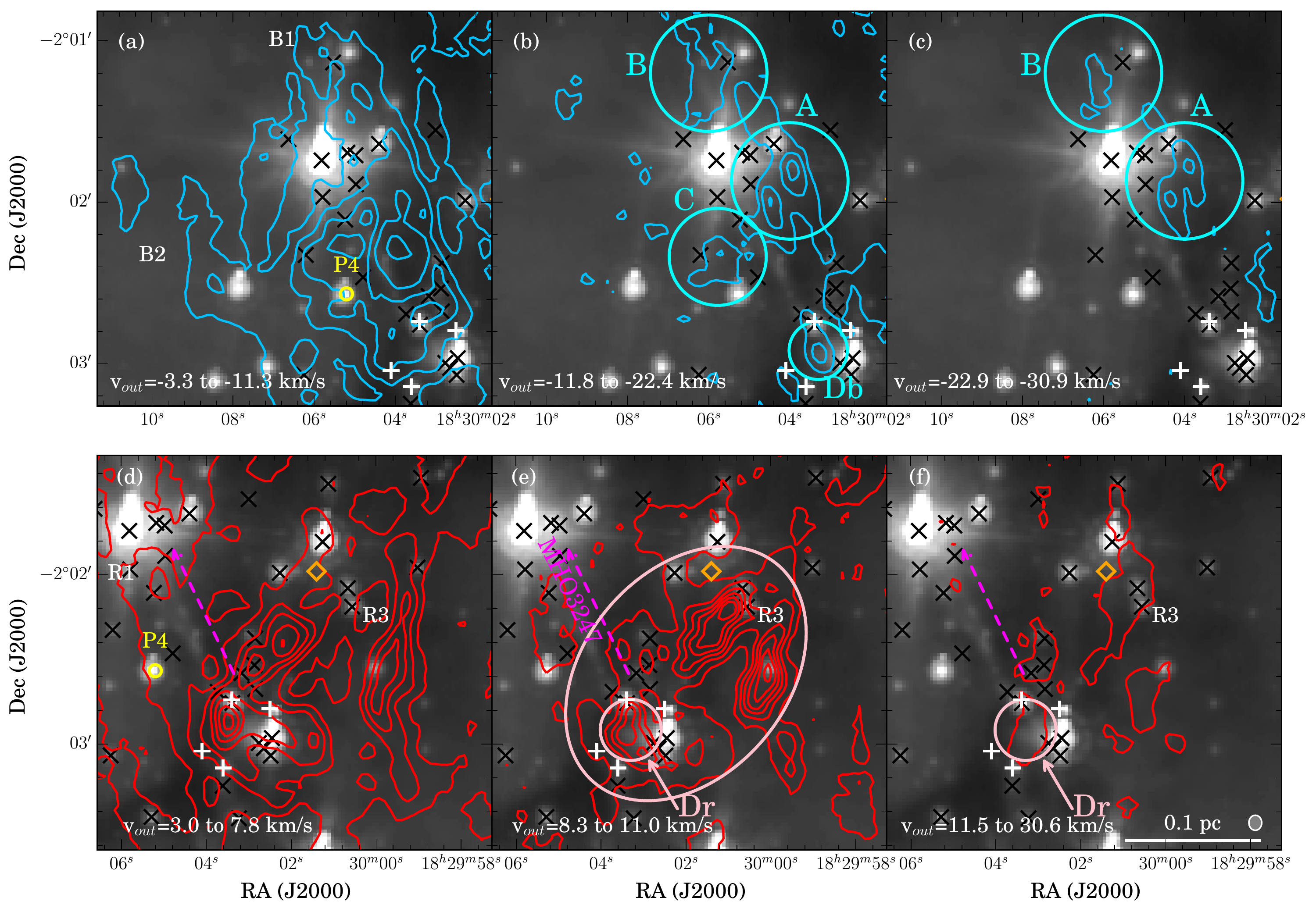}
\caption{$^{12}$CO outflow emission, showing low-velocity (left), mid-velocity (middle) and high-velocity (right) outflowing gas features within Box 1 (panels a-c) and Box 2 (d-f), respectively (see Figure \ref{fig:outflows}). Contours and background images are the same as in Figure \ref{fig:outflows}.  White plus signs show locations of continuum sources, while black crosses indicate \textit{Spitzer} protostars, as in Figure \ref{fig:carma_apex}. The beamsize and scale bar are given in the lower right corner.  Lines and ellipses indicate outflow features mentioned in the text. The dashed magenta line marking MHO 3247 replicates that of \citet[][see their Figure 1]{Tei12}, and coincides specifically to the 8$\mu$m emission feature slightly southeast of the line.  A, B, C, Db, and Dr are outflow features, and P4 is an IR source \citet{Tei12} that we describe in Section \ref{sec:results}.}  
\label{fig:outflowsz}
\end{figure*}

\begin{figure*}[!ht]
\includegraphics[height=0.41\textheight]{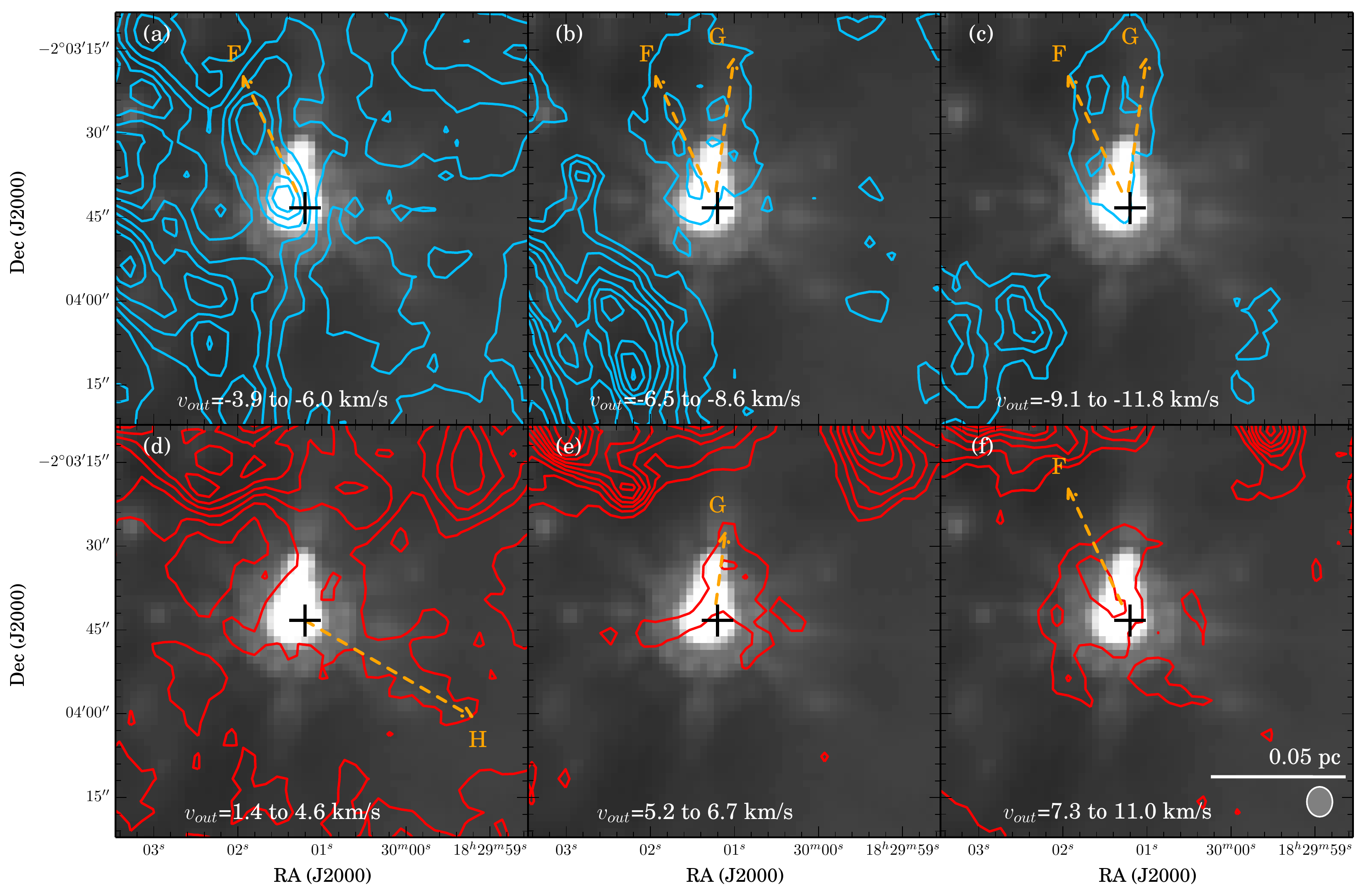}
\caption{$^{12}$CO emission near the candidate outflow from CARMA-1 (cross), corresponding to Box 3 in Figure \ref{fig:outflows}.  Contours are in units of rms for each moment map, beginning at a level of $4\sigma$ and incrementing by $4\sigma$ at low-velocity, and $3\sigma$ for other panels.  The beamsize and scale bar are given in the lower right corner.  Velocity intervals are indicated for each panel, where outflow velocity, $v_{out}$, is relative to the central cloud velocity of \vcloud. Features marked with letters and arrows are addressed in Section \ref{sec:carma1out}.}  
\label{fig:outflow_cand1}
\end{figure*}

Prevalent blueshifted CO features appear northeast and south of the continuum sources CARMA-3/CARMA-5 and CARMA-6/CARMA-7, shown in Figures \ref{fig:carma_apex}-\ref{fig:channel2} and the upper row of Figures \ref{fig:outflows}-\ref{fig:outflowsz}.  In the lower blue velocity channels (i.e., especially with approximately $v_{out} = -6$\ to $-4$\ \kms), these features appear as wide-angle outflows with opening angles of $\sim$70\dg\ and 45\dg, and position angles of $\sim$45\dg\ and 180\dg, respectively. However, they may in fact be several more collimated outflows super-imposed on one another (see Figures \ref{fig:channel2}, \ref{fig:outflows} and \ref{fig:outflowsz}).  These outflowing structures correspond approximately to those identified as B1, B2, B3 and B4 by \citet{Nak11}, which we have also labeled in Figure \ref{fig:outflows}(a) for reference.  The outflow features in low-velocity channels suggest that B1 and B2 comprise the cavity walls of a single northeastern outflow lobe, and similarly B3 and B4 comprise the walls of a southern outflow lobe.  Wide-angle, far-reaching SHV outflows with cavity walls that are nearly parallel at larger distances from the driving sources have been observed in other regions, including IRAS 2-south in NGC 1333 \citep{Plu13}, HH 46/47 \citep{Arc13}, and RNO 91 \citep{Lee05}.  In the same direction as the outflows B1 and B2 to the northeast, we see that the peaks of \thco\ emission correspond to the strongest \twco\ emission, shown in Figure \ref{fig:12co_13co}(a) (see also Figure \ref{fig:channel2b}), indicating entrained gas even at lower velocities than what is probed by the optically thick \twco.  We note also that since protostars in Serpens South are in such close proximity, and given that we expect each protostellar source to drive a bipolar outflow, it is likely that what appear as broad or wide-angle outflows are in fact comprised of several more collimated outflows from the distinct continuum sources mentioned here or from the other $\sim10-20$ protostars that reside in the central region but remain undetected in continuum observations. 

The northern wall of the blue northeast outflow structure is comprised of clumpy emission (features A and B in Figure \ref{fig:outflows}(b)-(c)) along an axis with position angle of $\sim20\dg$, appearing to originate from near the sources CARMA-3 and CARMA-5 and extend $\sim120\arcsec$ (0.25 pc) northeast.  At blue velocities with $|v_{out}|\gtrsim17$ \kms, the outflow is comprised of delineated clumpy features that form part of the structure's wall.  Another clumpy feature (C in Figure \ref{fig:outflows}(b)) is seen up to $|v_{out}|=15$\ \kms\ at a location 45\arcsec\ (0.1 pc) northeast of CARMA-3 and CARMA-5 (along an axis that intersects CARMA-5 with position angle $\sim 60\dg$).  Outflow emission in the northeast is seen with $|v_{out}|$ up to $\sim30$\ \kms\ (Figure \ref{fig:outflows}a-c), whereas the southern blue outflow shows emission up to $|v_{out}|\sim10$ \kms\ (Figure \ref{fig:outflows}(a)).  

These clumpy features may be evidence of episodic ejection from one or more sources in the central region, but we cannot be certain of whether these come from one or more driving sources in that region.  We also note that the clumpy and bullet-like blueshifted emission in high-velocity channels is sufficiently strong and localized that negative side lobes appear at their edges due to the synthesized beam, as seen in the upper left panels of Figure \ref{fig:channel2}.  These negative sidelobes have absolute values that are less than 4 sigma in individual channels, and they are not included in any calculations of mass, momentum or energy (see Section \ref{sec:mass}).

Several ``infrared H-H objects'' \citep[e.g., described by][]{Nak11} and H$_2$ features \citep{Tei12} may be associated with the CO outflows discussed here.  \citet{Nak11} report the infrared H-H objects, likely resulting from shocks by protostellar outflows, which they call K1 and K2 toward the southwest of the map.  These H-H objects have bow shapes indicating that they are moving toward the southwest and probably originated near the region of high protostellar density at the center of our map.  \citet{Tei12} detected two H$_2$ features (MHO3249b and MHO3250) in which the emission appears to correspond with the IR H-H objects K1 and K2, respectively \citep{Nak11}.  \citet{Tei12} suggest that P4 drives MHO3249b, Y1 (our CARMA-1) drives MHO3250 (described below), and P3 (our CARMA-6) drives H$_2$ features southward (beyond the edge of our map) including MHO3248 (see \S \ref{sec:continuumsources} for continuum source labels).  This implies that the western component of the southern blue outflow we detect is driven by CARMA-6 (P3) and the eastern component may be related to the source P4 from \citet{Tei12}. However, in this mapped region there are several plausible driving sources, and higher-resolution data are needed to determine their origin(s), as well as whether these components may form cavity walls of the same outflow.  Figure \ref{fig:outflows} marks these features from the previous literature with magenta dashed lines and ellipses.

One \twco\ emission feature that appears very compact and is apparent thanks to our high-resolution data is located approximately equidistant from the continuum sources CARMA-3, CARMA-5, CARMA-6 and CARMA-7 (feature Db in Figure \ref{fig:outflows}b), but not coincident with any single source detected here or identified in the literature to our knowledge.  Molecular CO emission is seen here in velocity channels up to $|v_{out}|=20$\ \kms.  This outflow's redshifted counterpart (feature Dr, discussed in the following section), is coincident (on the plane of the sky) with this blueshifted feature, and therefore this may be a bipolar outflow oriented nearly along the line of sight.  The compact morphology is similar to that of IRAS 4B in NGC 1333 \citep{Yil12,Plu13}.

\subsubsection{Redshifted outflows}  \label{sec:redout}

Most redshifted outflow emission is located north-northwest of the continuum sources in our map (see Figures \ref{fig:carma_apex}-\ref{fig:channel2} and lower panels of Figures \ref{fig:outflows}-\ref{fig:outflowsz}), comprising a structure with position angle of $-20\dg$ to $-30\dg$ and reaching about $210\arcsec$ (0.4 pc) from the continuum sources.  At the farthest extent of this emission, especially at low velocities with $|v_{out}|\lesssim8.5$ \kms, the emission has bow-shaped morphology, and at higher velocities the emission is more clumpy (labeled E in Figures \ref{fig:outflows}(e) and (f)).  This far-reaching redshifted emission appears to correspond to R2 in \citet{Nak11}, but the features are seen in more detail here.  Additionally, in Figure \ref{fig:12co_13co} we show the expansive \twco\ redshifted emission overlaid on the low-velocity \thco\ blueshifted emission. It appears that the dense gas traced by \thco\ emission at low-velocities is being entrained and spanning an even larger region than can be seen with \twco\ (due to its higher optical thickness), with especially strong \thco\ emission seen to the northwest and south. 

At velocities up to about  $|v_{out}|=12$\ \kms\ (shown in Figures \ref{fig:outflows}(d) and (e), coinciding with the label R3), \twco\ emission forms an oblong cavity that extends from the location of the continuum sources northwest about $\sim100\arcsec$ (0.21 pc).  It is not clear where this SHV emission originates, or whether it is the superposition of several outflows originating from distinct locations.  In low-velocity channels, the \thco\ emission corresponds to the peaks of \twco\ emission that outline this feature (see Figure \ref{fig:12co_13co}b).  At higher velocities, the \twco\ structure appears clumpier, while at lower velocities the structure is more widespread and diffuse.

At low red velocities (see Figure \ref{fig:outflows}(d)), we detect a very collimated feature with position angle $\sim$30\dg\ (i.e., northeast) that extends about 140\arcsec\ (0.28 pc) from the central cluster of continuum sources and reaches a velocity of $|v_{out}|=8$\ \kms. This feature is labeled as R1 by \citet{Nak11} and coincides with MHO3249r by \citet{Tei12}.  A line drawn through the feature with approximately the same position angle passes close to CARMA-6 and CARMA-7 to the southwest, suggesting that one of these sources (or another one of the $\sim10$ nearby protostars seen with \textit{Spitzer}) is a possible driving source.  Another, shorter emission feature to the west/southwest of MHO3249r and corresponding to MHO3247 \citep{Tei12}, appears to originate from near CARMA-5 and CARMA-3 and reaches a distance of 60\arcsec (0.12 pc) northeast of CARMA-5 (see magenta annotation in Figures \ref{fig:outflowsz}(d)-(f)).  The position angle of this feature is slightly less than that of the longer collimated flow previously mentioned in this paragraph, and it is also possible that these two features form parts of the same flow, in which case the flow may be precessing.

A high-intensity clump is seen equidistant from the continuum sources and coincident with the blueshifted clump previously mentioned, with redshifted outflow emission up to $|v_{out}|=18$\ \kms\ (marked as Dr in Figures \ref{fig:outflows}(e) and (f)).  Since we detect emission at this location in blue and redshifted velocity channels with approximately $v_{out}=\pm20$ \kms, this appears to be an outflow oriented along the line of sight (see Figures \ref{fig:outflowsz}(d)-(f).  Figure \ref{fig:outflows}(b) marks the blueshifted counterpart Db at the corresponding location on the plane of the sky.  Although we do not confirm a driving source with continuum emission, this emission is nearby several MIPS 24 $\mu$m sources, and hence the outflow may be associated with a source that is very embedded.  

\subsubsection{Outflow emission near CARMA-1}  \label{sec:carma1out}
Given the relative isolation of CARMA-1 in this field (compared with the sources in the center of the cluster, including CARMA-3, CARMA-5, CARMA-6, and CARMA-7), we made a close channel-by-channel inspection of emission in the immediate vicinity of CARMA-1 to identify outflows uniquely associated with this Class I source.  This region is shown in more detail in Figure \ref{fig:outflow_cand1}, and we highlight the low-velocity channels where outflow morphology is apparent. Previously, this source \citep[labeled as Y1 by][]{Tei12} was proposed to drive the H$_2$ feature MHO3250 toward the southwest, and \citet{Nak11} labeled two outflow features (R4 and B5) that were possibly driven in that direction by this source.  Qualitatively our data support the previously proposed source and outflow identifications, but now we detect emission closer to the source, and it suggests that there may be additional outflow components associated with this source. 

Several clumpy \twco\ redshifted features appear to be concentrated near CARMA-1, shown in Figures \ref{fig:outflow_cand1} (d)-(f), with features typically appearing in only several contiguous velocity channels.  For example, at redshifted velocities $|v_{out}|<5$\ \kms, a narrow feature extends southwest $\sim 35\arcsec$ with a position angle of $240\dg$; this feature is oriented along the same direction as the feature R4 of \citet{Nak11} previously mentioned.  Additionally, blueshifted emission is seen super-imposed in low-velocity channels ($|v_{out}|<6$\ \kms, Figure \ref{fig:outflow_cand1}(a)), which entails that these features trace the cavity walls and this outflow is approximately in the plane of the sky.  The low radial-velocities that we detect here and the proposed orientation in the plane of the sky are consistent with the related MHO being located such that the outflow has a relatively large spatial extent ($\sim90\as$).  However, the blueshifted emission feature is oriented slightly more south than the redshifted feature, and therefore they may not necessarily form the same coherent structure.

At blueshifted velocities with $|v_{out}|<10$\ \kms\ we detect clumpy emission with position angle $\sim25\dg$ and extending $\sim30\arcsec$ (0.06 pc) northeast from CARMA-1, marked as feature F in Figure \ref{fig:outflow_cand1}(a)-(c).  Emission in this direction is also seen super-imposed in redshifted velocities with $|v_{out}|\gtrsim7$\ \kms\ (Figure \ref{fig:outflow_cand1}f).  At slightly higher velocities -- seen particularly in Figure \ref{fig:outflow_cand1}(b)-(c) and marked as feature G -- blueshifted molecular outflow emission is seen to extend $\sim35\arcsec$ north of CARMA-1, with position angle of $0\dg$ to $-10\dg$.  It is not clear whether the components with position angles of $\sim25\dg$ and $\sim-10\dg$ to the north of CARMA-1 correspond to the same outflow with opening angle of $35\dg$, or whether these are distinct outflows emanating from multiple sources in the proximity.  Also, in redshifted velocities shown in Figure \ref{fig:outflow_cand1}(e) emission is seen to the north of CARMA-1 with a similar position angle, perhaps constituting the far-side wall of the same outflow and suggesting that the inclination of this outflow is nearly in the plane of the sky. In the \thco\ map, we inspected low-velocity channels in the vicinity of CARMA-1 and we found \thco\ emission, in particular, to the north of CARMA-1 (see Figure \ref{fig:12co_13co}), although the emission is somewhat diffuse and does not conclusively point to this source. 

In summary, here we provide additional evidence that one or more molecular outflows are being driving from the location of CARMA-1.  However, a clear picture of a bipolar outflow is not immediately apparent.  Further, the collimated low-velocity blueshifted feature F which is a candidate outflow lobe does not have an equally collimated counterpart in the red.  It seems that the outflow structure of this source is either irregular, or the emission is being overwhelmed by stronger emission from the surrounding region which makes the outflow appear more clumpy rather than collimated.  Alternatively, if gas has already been swept away from the immediate region by other stronger outflows such as those to the northeast, there remains little gas to be entrained by this driving source CARMA-1.

\begin{deluxetable*}{lccccc}
\tablecaption{Outflow mass, momentum, energy \label{tab:mass}}
\tabletypesize{\small}
\setlength{\tabcolsep}{0.04in} 
\tablehead{\colhead{Method} & \multicolumn{3}{c}{Mass} & \colhead{Momentum} & \colhead{Energy} \\ 
\cline{2-4}
\colhead{}&\colhead{All (M$_\odot$)}  &  \colhead{Blue\tablenotemark{a} (M$_\odot$)} & \colhead{Red\tablenotemark{b} (M$_\odot$)} & \colhead{(M$_\odot$ km s$^{-1}$)} & \colhead{($\times10^{45}$\ erg)} } 
\startdata
(1) $\tex(x,y,v)$, pixel-by-pixel + channel-by-channel \tablenotemark{c} & 2.8 & 1.2 & 1.6 &  13.6 & 1.0 \\ 
(2) $\tex(v)$, channel-by-channel \tablenotemark{c} & 2.9 & 1.2 & 1.7 &  13.9 & 1.0 \\ 
(3) T$_{ex}$ = 10 K & 2.2 & 1.1 & 1.2 &  10.8 & 0.8 \\ 
(4) T$_{ex}$ = 30 K & 4.6 & 2.3 & 2.4 &  22.4 & 1.7 \\ 
(5) T$_{ex}$ = 50 K & 7.2 & 3.5 & 3.7 &  34.7 & 2.6 \\ 
\enddata
\tablenotetext{a}{Blue channels correspond to $|v_{out}|$ = 3.3 to 30.9 \kms.}
\tablenotetext{b}{Red channels correspond to $|v_{out}|$ = 3.0 to 30.6 \kms.}
\tablenotetext{c}{Means: $\overline{\tex}(x,y,v)=17\pm8$\ K for (1), and $\overline{\tex}(v)=15\pm4$\ K for (2).} \todo{Value}
\end{deluxetable*}

\subsection{Mass, momentum and energy of outflowing gas} \label{sec:mass}

The mass, momentum and energy of outflowing gas were calculated in order to assess the overall contribution of outflow activity within the mapped region, and we compared these with turbulent momentum and energy and gravitational potential energy.  First, we calculated H$_2$ column density ($N_{H_2}$) for each pixel ($x,y$) and velocity channel (with channel width $\Delta v$) as a function of excitation temperature (\tex) according to the following relations \citep[following][]{Dun14}: 
\begin{eqnarray}
N_{H_2}(x,y,v,\tex)& = &f(J,\tex,X_{CO}) T_{mb,12}^* \Delta v \label{eq:NH2}\\ 
f(J,\tex,X_{CO})&=&X_{CO,12}^{-1}\frac{3k}{8\pi^3\nu\mu^2}\frac{2J+1}{J+1}\frac{Q(\tex)}{g_L}e^\frac{E_{J+1}}{k T_{ex}}. \label{eq:f}
\end{eqnarray}
The opacity correction is explained in Appendix \ref{sec:opacity}, with the main-beam temperature of \twco\ corrected for optical depth, $T_{mb,12}^*$, given by Equation \ref{eq:tmbcorr}.  Appendix \ref{sec:tex} describes the method(s) to solve for excitation temperature, \tex.  In the above equations, $J=0$ is the lower energy level of the rotational transition, and $g_J=2J+1=1$ is the degeneracy of the lower state. For $X_{CO}$ we assume a standard abundance ratio $[^{12}\textrm{CO}/\textrm{H}_2]=10^{-4}$ \citep{Fre82}.  For \twco\ $J=1-0$, $\nu=115.27$\ GHz, and $\mu$ is the \twco\ dipole moment with a value 0.11 Debye.  $Q(\tex)$ is the partition function which here is approximated as $Q(\tex)=k\tex/{hB}$, and $B=57.635$ GHz is the rotation constant for CO.  $E_{J+1}=7.64\times10^{-16}$ erg is the energy of the upper level such that the energy of the transition is $E=E_{J+1}-E_J$, and $E_{J+1}/k=5.53$ K.  

Having measured column density for each pixel, mass is then determined to be:
\begin{eqnarray}
M(x,y,v)& = &2.72 m_H N_{H_2}(x,y,v) A,
\end{eqnarray}
where $m_H$ is the mass of a hydrogen atom, such that 2.72$m_H$ is the mean molecular weight of the gas \citep{Bou97}, and $A$ is the area of a pixel.  We sum over the cube in spatial and spectral dimensions including pixels that have signal greater than 3$\sigma$ to get the total outflow mass, momentum and energy:

\begin{eqnarray}
M_{out}& = &\Sigma_{area} \Sigma_{vel} M(x,y,v) \\
P_{out}& = &\Sigma_{area} \Sigma_{vel} M(x,y,v) |v-v_{cloud}| \\
E_{out}& = &\frac{1}{2}\Sigma_{area} \Sigma_{vel} M(x,y,v) |v-v_{cloud}|^2, \label{eq:energy}
\end{eqnarray}
where $v$ is the velocity corresponding to a given channel, and $v_{cloud}=\vcloud$ is the cloud velocity.  Mass, momentum and energy for velocity channels with $3 < |v_{out}| < 31\ \kms$ and the region we mapped covering $\sim45$ arcmin$^2$ (see Figure \ref{fig:cover}) are given in Table \ref{tab:mass}.  The lower-limit of this velocity range was chosen by inspecting low-velocity channels by eye to exclude those channels dominated by ambient cloud emission, and the upper-limit was dictated by the spectral coverage of our observations.  We measure a total outflow mass of 2.8 \msun, of which blueshifted and redshifted channels contribute 43\% and 57\%, respectively.   \todo{Value} These masses were calculated using the $\tex(x,y,v)$ values determined with the pixel-by-pixel + channel-by-channel (velocity-dependent) method as described in Appendix \ref{sec:tex} (a.k.a. method 1).  Using excitation temperatures determined according to the alternative methods 2-5 (these methods either spatially average over pixels in each channel, or assume a constant excitation temperature, as described in Appendix \ref{sec:tex}) result in total outflow masses ranging from 2.2 to 7.2 \msun, listed in Table \ref{tab:mass}.   \todo{Value} 

\begin{figure*}[!ht]
\includegraphics[width=\linewidth,angle=0]{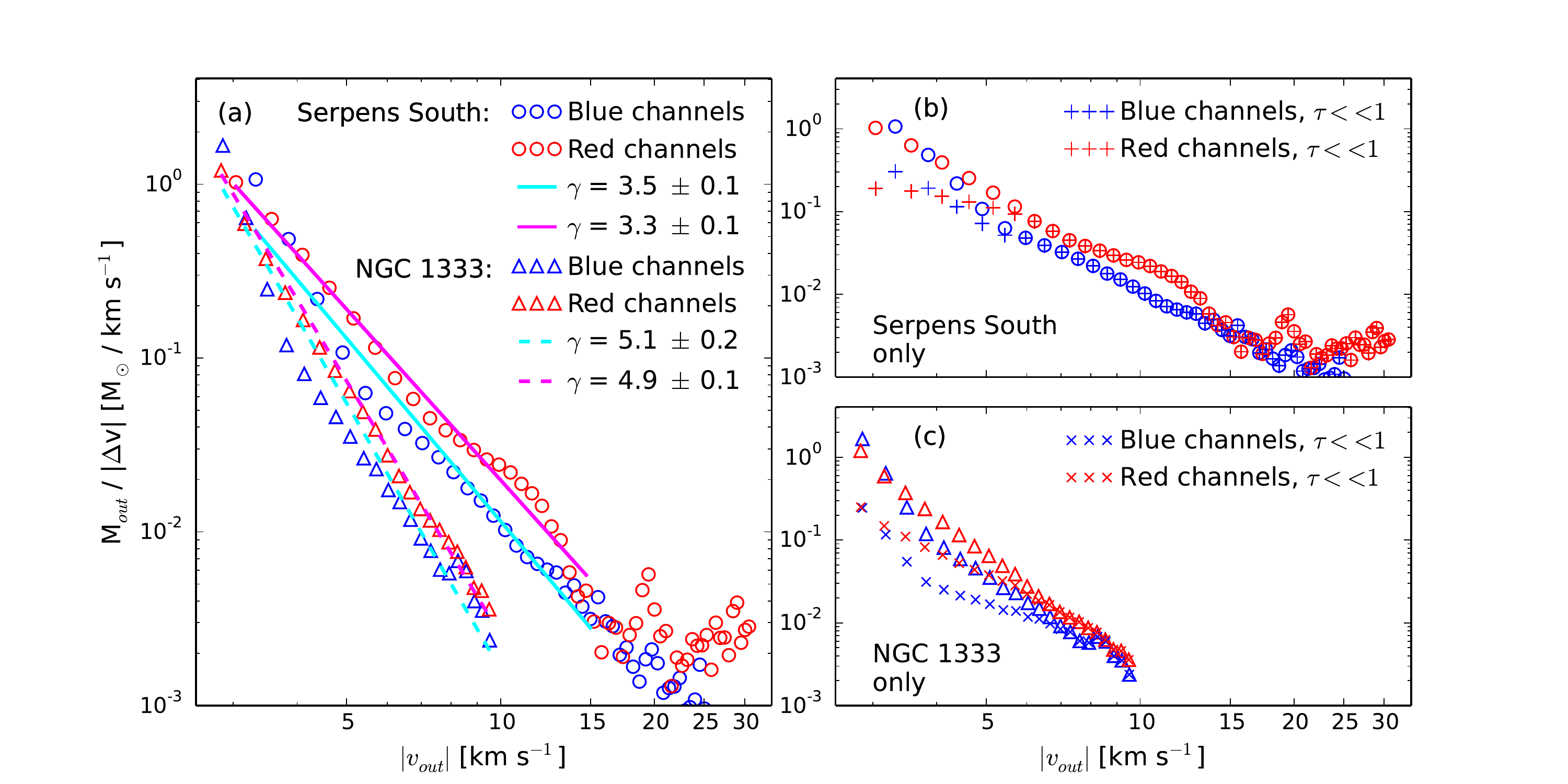}
\caption{Outflow mass spectra for Serpens South (circles) and NGC 1333 (triangles), with blue and redshifted velocity channels plotted independently with their respective colors.  Panel (a) shows both regions for comparison, and aqua and magenta lines (solid for Serpens South, dashed for NGC 1333) are the best-fit power-laws to the blue and redshifted mass spectra, respectively.  Panels (b) and (c) show Serpens South and NGC 1333 separately, where mass spectra with no velocity-dependent correction for opacity (i.e., assuming $\tau<<1$) are also given by plus and cross symbols for Serpens South and NGC 1333, respectively. }  
\label{fig:mass_spectra}
\end{figure*}

In Figure \ref{fig:mass_spectra} we show the outflow mass distribution per velocity bin for Serpens South (panels a and b).  For the blue and redshifted mass spectra, we fit power laws of the form $m(v) \propto v^{-\gamma}$ for the velocity range $3 < |v_{out}|< 15\ \kms$, and we found slopes of $\gamma=3.5\pm0.1$\ and $\gamma=3.3\pm0.1$, respectively\todo{Value}.  The slope may be an evolutionary indicator of mass ejection in a cloud, such that the slope steepens with time \citep{Smi97}.  We performed a similar analysis of the outflow mass within the slightly more evolved region NGC 1333 \citep{Plu13}, assuming $\tex=17$\ K which is the mean \tex\ in Serpens South, and the blue and redshifted mass spectra for that region can be fit with power laws with slopes of $\gamma=5.1\pm0.2$\ and $\gamma=4.9\pm0.1$ (see Figure \ref{fig:mass_spectra}, panel a and c).  These two regions are therefore presented as a examples for which the shallower (steeper) slope pertains to the less (more) evolved region.

It has been previously suggested that a break in $\gamma$ at $\sim 10\ \kms$ can be interpreted as two distinct outflow velocity components \citep[e.g.,][]{Ric00,Su04}.  In that case, the high-velocity gas corresponds to a recently accelerated component and steeper slope ($\gamma \sim 3-4$), and the low-velocity gas is a slower, coasting component with a less steep slope ($\gamma\sim2$).  Neither Serpens South nor NGC 1333 showed a break in their power spectra according to our primary analysis.  We note that our analysis included a velocity-dependent correction for opacity of the \twco\ line, which especially accounted for more ``hidden'' mass at low-velocities than in the case of the optically thin assumption.  When assuming that the \twco\ line is optically thin ($\tau<<1$), breaks in the power spectra are apparent at outflow velocities of $\sim 5-10$\ \kms\ (see Figure \ref{fig:mass_spectra} b-c).  Our findings are in line with the suggestion by \citet{Bal99} that a broken power-law index for outflows may not be a physical distinction, but rather an artifact of the optically thin assumption. 

Comparing blue and redshifted mass at each outflow velocity, the ratio $M_{out,blue}/M_{out,red}(|v_{out}|)$ ranges from 0.1 to 1.5, with a mean of 0.5\todo{Value}.  While we expect outflows to be bipolar, and hence the blue and redshifted gas should total approximately equal amounts, in this map it appears that slightly more gas is being entrained by redshifted lobes than by blueshifted lobes, especially in velocity channels with $|v_{out}|\sim5-10\ \kms$.  Possibly some outflow gas has escaped the cloud more readily in the foreground than in the background, or extends beyond the map edges, depending on the locations of the driving protostars within their nascent gas cloud.  This mass ratio may also entail that the surrounding gas environment was not homogeneous in mass and/or density at the time when protostars began to drive outflows.  We present maps of \tex\ in Appendix \ref{sec:tex}, and the highest \tex\ values are found in redshifted velocity channels.

In addition to the mass calculation, we calculate outflow momentum to be $P_{out}=13.6\ \msun\ \kms$, and energy to be $E_{out}\sim1.0\times10^{45}$\ erg\todo{Value}.  Table \ref{tab:mass} lists these diagnostics using each \tex\ method, and we consider these values to be lower limits because they are uncorrected for inclination angle of the outflows.  Assuming an average inclination angle with respect to the line of sight $\xi=57.3\dg$\ \citep[e.g.,][]{Nak11,Dun14}, and assuming that all outflow motion is along the jet axis \citep[see][]{Dow07}, results in greater momentum and energy by factors of 1.9 and 3.4, respectively.  Hence, momentum and energy from method 1 are  $P_{out}=25.3\ \msun\ \kms$ and $E_{out}=3.3\times10^{45}$\ erg, after correction for an average inclination angle of $\xi=57.3\dg$.  \todo{Value}

We performed a comparison of the cumulative mass and energy of outflows with those determined by \citet{Nak11} using \twco\ $J=3-2$\, assuming: (1) constant \tex=30 K, (2) distance of 260 pc, and (3) including only the velocity channels between $-15\ \kms$\ to $4\ \kms$, and $11$\ to $30\ \kms$, with a cloud velocity $v_c=7.5\ \kms$.  We found the cumulative mass and energy of outflows to be 1.8 and 1.1 times the values presented by \citet{Nak11}, respectively.  In addition to using a different energy level transition, this discrepancy may also be attributable to the correction we make for opacity of CO, whereas \citet{Nak11} assume \twco\ $J=3-2$\ is optically thin.  On the other hand, we also note that the map coverages are not identical, and specifically the blue features B7, B8, and B9 from \citet{Nak11} are not covered entirely by our map.  Nonetheless, we consider these masses and energies to agree within the uncertainties that we cannot account for here.

As previously mentioned (see Section \ref{sec:mass} for more discussion), we have assumed a minimum outflow velocity of $|v_{out}|=3\ \kms$, so as not to confuse emission from the ambient cloud.  However, as \citet{Dun14} suggest, typical escape velocities from individual sources are less than this, and therefore our mass calculation potentially omits some outflowing gas.  Making an admittedly simple extrapolation to velocities as low as $v_{out}=0.3\ \kms$, based on the spectral fit shown in Figure \ref{fig:mass_spectra}, we find that the total mass may increase by an order of magnitude or more, which was also shown by \citet{Dow07, Off11, Dun14} for low-velocity outflows. When assessing the impact that outflows have on the cloud environment, the momentum and energy are more appropriate diagnostics than mass, as they will be less affected by the low-velocity channels (i.e., a factor of a few when extrapolating based on Figure \ref{fig:mass_spectra}) so that the main conclusions that we propose below will hold qualitatively.

\begin{deluxetable*}{lccccccl}
\tablecaption{Outflow Properties \label{tab:outdiag}}
\tabletypesize{\footnotesize}
\setlength{\tabcolsep}{0.04in} 
\tablehead{\colhead{Property} & \colhead{Inclination} & \multicolumn{2}{c}{Limits\tablenotemark{a}}  && \multicolumn{2}{c}{Medians\tablenotemark{b}}&\colhead{Units}  \\ 
\cline{3-4}  \cline{6-7}
\colhead{} & \colhead{Dependence}  &  \colhead{Lower} & \colhead{Upper} && \colhead{Uncorrected} & \colhead{$\langle\xi\rangle=57.3\dg$}&\colhead{} }
\startdata
Characteristic length ($R_{char,out}$)  & $1/\sin\xi$&0.1 & 0.4 && 0.19 &  0.22& [pc] \\ 
Characteristic velocity ($V_{char,out}$) &  $1/\cos\xi$ & \nodata &\nodata && 4.8 &  8.9 & [\kms] \\ 
Dynamical time ($t_{dyn}$)& $\cos\xi/\sin\xi$  &1.9 & 8.9 && 3.8 &  2.4 & [$\times10^4$\ yr ] \\ 
Mass loss rate ($\dot{M}$)& $\sin\xi/\cos\xi$&1.5 & 0.3 && 0.7 &  1.2 & [$\times10^{-4}\ \msun$\ yr$^{-1}$] \\ 
Outflow force ($F_{out}$)& $\sin\xi/\cos^2\xi$ &7.2 & 1.5 && 3.6 &  10.3 & [$\times10^{-4}\ \msun$\ \kms\ yr$^{-1}$]\\ 
Outflow luminosity ($L_{out}$)& $\sin\xi/\cos^3\xi$&42.2 & 9.0 && 21.1 &  112.6 & [$\times10^{-2} L_\odot$] \\ 
\enddata
\tablenotetext{a}{``Lower'' and ``upper'' correspond to outflow property listed in first column obtained using the lower and upper limits of $R_{char}$ (row 1), respectively, and uncorrected for inclination angle.}
\tablenotetext{b}{Correspond to the outflow properties listed in first column obtained using the median $R_{char}$ value (row 1), with ``uncorrected" indicating raw values without correcting for inclination angle, and $\langle\xi\rangle=57.3\dg$ is the assumed average outflow inclination angle with respect to the line of sight.}
\end{deluxetable*}

Conversely, the high-velocity outflows contribute more significantly to the momentum and energy budgets, rather than mass.  The velocity range that we probe here, with $|v_{out}|<31\ \kms$, is mostly representative of the SHV outflow component.  However, EHV components have been shown to be driven by some low-mass protostars \citep[e.g.,][]{Taf04}, and if present they can account for 2-4 times as much momentum and energy as the SHV components.  In the following sections we compare the relative contributions of momentum and energy from outflows, turbulence, and gravity of the cloud.  Although the magnitude of the relative contribution by outflows may change if in fact there are EHV components associated with these sources, our discussion and conclusions (below) that outflows significantly impact the surrounding environment hold.  Nonetheless, given that we do not probe the entire range of low- and high-velocity outflow gas, it should be kept in mind that we likely under-estimate the mass and subsequent mass-dependent kinematic properties expressed in Sections \ref{sec:prop}-\ref{sec:gravity}.

\section{Discussion} \label{sec:discussion}

\subsection{Characteristic outflow properties} \label{sec:prop}

In order to characterize the outflows in this region, we present several common outflow properties in Table \ref{tab:outdiag}.  We list the dependence of each property on assumed inclination in column 2, where $\xi$ is the angle between the outflow axis and the line of sight (i.e., $\xi=90\dg$\ corresponds to an edge-on system).  For each property, we present values corresponding to the lower and upper limits of the characteristic length (columns 3-4); we also calculate the values corresponding to the median characteristic length or dynamical time, both uncorrected for inclination angle (column 5) and assuming an average inclination angle of $\langle\xi\rangle=57.3\dg$\ (column 6).  The respective units are shown in column 7.

The first outflow property is the characteristic outflow length ($R_{char,out}$), and this value propagates in subsequent calculations.  In Section \ref{sec:outflows}, we discussed several coherent outflow features which have lengths of $45\arcsec - 210\arcsec$\ ($R_{char,out}=0.1$\ pc to 0.4 pc, respectively), and a median of $90\arcsec$ ($R_{char,out}=0.19$\ pc, or 0.22 pc for $\xi=57.3\dg$). \todo{Value}  Based on the morphology of the emission detected here, we suggest that the majority of the outflows that we discussed originate from the YSOs near the dense cluster center in the map, and we assume that the previously mentioned values for $R_{char,out}$ are the limits and median lengths representative of outflows found in this region.  We use these values in further diagnostic calculations, although it has also been shown that outflow lengths vary by at least an order of magnitude in other regions \citep[e.g.,][]{Nak11Oph,Plu13}, so more extreme values may be feasible.  Nonetheless, our median value and range that we present in Table \ref{tab:outdiag} are consistent with the literature, especially \citet{Nak11} who estimated the mean length of outflow lobes in Serpens South to be $0.1-0.2$\ pc, or $0.17-0.33$\ pc at the assumed distance of 429 pc.  \todo{Value} 

Next, we calculated the characteristic velocity ($V_{char,out}$) which is the total outflow momentum divided by outflow mass in the region, and we found that  $V_{char,out}=P_{out}/M_{out}=4.8\ \kms$. \todo{Value} Correcting for an inclination angle $\xi=57.3\dg$ resulted in $V_{char,out}=8.9\ \kms$\todo{Value}.  The resulting median dynamical time defined as $t_{dyn} = R_{char,out}/V_{char,out}$\ is $3.8\times10^4$\ yr ($2.4\times10^4$\ yr for  $\xi=57.3\dg$), with a range from $1.9 - 8.9 \times10^4$ yr for the $R_{char,out}$ range mentioned previously\todo{Value}.  The remainder of diagnostics in Table \ref{tab:outdiag} pertain to outflow mass, momentum, and energy, and we explain those diagnostics in the following sections.

\subsection{Outflow momentum and energy compared with turbulence} \label{sec:turbulence}

In the model of \citet{Nak07}, outflows balance infall motions and accretion, sustain the cloud in quasi-static equilibrium, and lead to a slow SFR.  Without outflows to inject additional turbulent motions in this model, the turbulence would dissipate in less than several free-fall times.  Therefore, the following calculations aim to quantify momentum and energy injection by outflows, specifically in the context of infall and turbulence in the same region.  The calculations of mass loss rate, outflow force, and luminosity (see rows 4-6 of Table \ref{tab:outdiag}) are inversely proportional to the dynamical time, which is a source of uncertainty of a factor of a few as explained in the previous section.  We report the median values in the following text, but Table \ref{tab:outdiag} also includes a range that corresponds to the lower and upper limits for $t_{dyn}$ (uncorrected for inclination angle), as well as median values, both uncorrected and corrected for an average inclination angle of $\langle\xi\rangle=57.3\dg$.

\citet{Rei01} explained that the ``unified'' outflow model relates outflows with jets and/or winds, such that the same primary engine powers a YSO's high-velocity jet/wind as well as lower-velocity lobes of an outflow traced by CO emission. Hence, the cumulative mass loss rate via outflows and the mass loss rate via winds can be expressed according to $\dot{M}_{out}V_{char,out}=\dot{M}_{wind}V_{wind}$.  From our CO observations, we found the cumulative mass loss rate via molecular outflows ($\dot{M}_{out}=M_{out}/t_{dyn}$) to be $1.2\times10^{-4}\ \msun$\ yr$^{-1}$ (assuming $\xi=57.3\dg$)\todo{Value}.  It follows that the cumulative wind mass loss rate is $\dot{M}_{wind}=1.0\times10^{-5}\ \msun$\ yr$^{-1}$, assuming $V_{wind}=100$\ \kms\ \citep[with a likely range of $V_{wind}=100-200$\ \kms, c.f.][and references therein]{Bal07,Arc07}.  Further, it has been shown that mass loss rates are tied to accretion rates \citep{Har00}.  Based on the wind mass loss rate, and assuming the fraction of stellar mass launched in a wind to be $f_w\sim0.1-0.3$ \citep{Shu88,Pel92,Lee07a,Lee07b}, we expect the total mass accretion rate onto the stars powering outflows in this region to be $\sim0.3-1.0\times10^{-4}\ \msun$\ yr$^{-1}$ (or lower by a factor of 2 if we assume the larger $V_{wind}=200$\ \kms).

\citet{Kir13} estimated the infall rate of material onto the cluster-forming clump to be about $\dot{M}_{infall}\sim2.2\times10^{-4}\ \msun$\ yr$^{-1}$ (assuming filament inclination angle that they find to be 12\dg, and scaling to a distance of 429 pc).  We can relate this quantity with the total accretion rate onto stars within the cluster using $\dot{M}_{acc}=f_{SFE}\dot{M}_{infall}$, with $f_{SFE}$\ typically in the range of 0.1-0.3 \citep{Lad03}, since not all the material that is added to the clump through gravitational infall will be used to make protostars (and disks).  Assuming $f_w=0.1$ and $f_{SFE}=0.3$, we then estimated the infall rate for this region to be $\sim3\times10^{-4}\ \msun$\ yr$^{-1}$, which is comparable to the infall rate onto the cluster found by \citet{Kir13}. Hence, the cumulative mass loss rate via molecular outflows that we measured and the observed infall rate appear to be consistent in Serpens South.

In addition to assessing the outflowing mass over time, we calculated the total outflow momentum per unit time, known as the cumulative outflow force ($F_{out}=P_{out}/t_{dyn}$), to be $10.3\times10^{-4}\ \msun$\ \kms\ yr$^{-1}$ (assuming $\xi=57.3\dg$). Further, the cumulative outflow luminosity ($L_{out}=E_{out}/t_{dyn}$) is $1.13\ L_\odot$. For comparison, \citet{Nak11} reported a range $L_{out}=0.8-1.7\ L_\odot$ (corrected for a distance of 429 pc), assuming an inclination angle of $\xi=57.3\dg$ and a range of characteristic lengths $R_{out}=0.1-0.2$\ pc\todo{Value}.  

Comparing the turbulence energy dissipation rate and the outflow luminosity quantifies the contribution of outflows and their impact on the cloud turbulence.  In order to determine this, first we solved for turbulent energy, which is defined as $E_{turb}=\frac{1}{2}M_{cl} \sigma_{3D}^2$.  Throughout, we adopted a cloud mass of $M_{cl}=220\ \msun$\ and a radius of $R_{cl}=0.17$\ pc, based on observations of HCO$^{+}$ ($J=4-3$) by \citet{Nak11} and adjusting for a distance of 429 pc; the mass is consistent within $\sim20\%$ of that estimated by \citet{Kir13} for the central cluster.  Also, 3D and 1D velocity dispersion were related according to the following relations:  $\sigma_{3D}=\sqrt{3}\sigma_v$\ and $\sigma_v=\Delta v_{turb}/(2\sqrt{2\ln 2})$. The velocity dispersion $\Delta v_{turb}=1.36\ \kms$ (Fern{\'a}ndez-L{\'o}pez, 2014, private communication) was determined from the FWHM of N$_2$H$^+$ which, traces mostly dense and quiescent gas. The resulting turbulent energy within the mapped region was found to be $2.2\times10^{45}$\ erg. \todo{Value}

The turbulence energy dissipation rate is given by $L_{turb}=E_{turb}/t_{diss}$.  The dissipation time is defined as $t_{diss}=t_{ff} (3.9 \kappa/M_{rms}) $\ \citep[c.f.][]{Mac99}, where free-fall time is $t_{ff}=\sqrt{3\pi/(32 G \rho)}$, and $\kappa=\lambda_d/\lambda_J\sim1$ assuming that the turbulence driving length ($\lambda_d$) is equal to the Jean's length ($\lambda_J$) of the clump.  We determined the gas density of the cloud ($\rho=3 M_{cl}/4\pi R_{cl}^3$) to be $\rho=8\times10^{-19}$\ g cm$^{-3}$\todo{Value}. The Mach number is $M_{rms}=\sigma_{3D}\sqrt{2.72m_H/(kT)}=5.5$, with $T=10.8$\ K \citep{Fri13}.  The resulting turbulent energy dissipation rate is $L_{turb}=0.3\ L_\odot$. \todo{Value} The ratio $r_L=L_{out}/L_{turb}\sim0.6-3$ (the range corresponds to $L_{out}$\ uncorrected for inclination angle, and assuming $\xi=57.3\dg$)\todo{Value}\ quantifies the potential for outflows to counteract turbulence dissipation in the region.  A value of at least 1 indicates that outflows drive energy comparable to or greater than the turbulent energy in the region, as is the case here after correcting for inclination angle.

Another way to assess the potential impact of outflows on the cloud turbulence is to compare the total outflow force and the turbulence momentum dissipation rate.  In the outflow-regulated cluster formation model, the outflow force should be comparable to or greater than the turbulence momentum dissipation rate in order for outflow feedback to replenish internal supersonic turbulence and maintain the clump close to quasi-virial equilibrium \citep{Li06,Nak14}.  Turbulence momentum dissipation rate of the internal turbulent motion is defined as $dP_{turb}/dt=-0.21 M_{cl} \sigma_{3D}/t_{diss}$ \citep[equation 4 of][]{Nak14}.  We found $dP_{turb}/dt=-2.8\times10^{-4}\ \msun\ \kms\ \textrm{yr}^{-1}$\todo{Value}, which is approximately 25\% \todo{Value} of the median outflow force (assuming $\xi=57.3\dg$).  Even accounting for a factor of a few uncertainty in the outflow force that we mention above, the ratio of  $dP_{turb}/dt$\ to $F_{out}$\ likely remains less than 1.  Assuming that outflows can transfer momentum and energy to a substantial fraction of the gas in the immediate vicinity of the dense cluster of young protostars, the ratio of outflow force to turbulence momentum dissipation along with the ratio $r_L$ (see above) suggest that outflows are an important source of turbulence in this region.

\begin{deluxetable*}{lccl}
\tablecaption{Comparison of Serpens South and NGC 1333 \label{tab:compare}}
\tabletypesize{\tiny}
\setlength{\tabcolsep}{0.02in} 
\tablehead{\colhead{Property} & \colhead{Serpens South \tablenotemark{a}}&\colhead{NGC 1333 \tablenotemark{a}}&\colhead{Units}  }
\startdata
Protostellar number density  & 130 & 110 & [pc$^{-2}$]\\
Protostellar fraction  & 80 & 50 & [\%]\\

Outflow mass per area  & 9.7 & 26& [$\msun$ pc$^{-2}$] \\
Outflow emission prevalence\tablenotemark{b}& 90 & 60& [\%] \\

$r_L\ (=L_{out}/L_{turb})$  & $0.6-3$\tablenotemark{c}  & $6-24$\tablenotemark{c} &\\
$\eta_{out}\ (=-2E_{out}/W)$  & $0.1-0.3$\tablenotemark{c} & $0.2-0.7$\tablenotemark{c} &\\
\enddata
\tablenotetext{a}{Pertains to area enclosed in white rectangle in Figure \ref{fig:continuum} for Serpens South, and region mapped by \citet{Plu13} for NGC 1333.}
\tablenotetext{b}{Outflow emission coverage (pixels projected onto the plane of the sky) with respect to a mapped area of 0.3 pc$^2$.}
\tablenotetext{c}{The given lower- and upper-limits of ranges correspond to (1) raw values without correcting for inclination angle, and (2) assuming $\langle\xi\rangle=57.3\dg$ as the average outflow inclination angle with respect to the line of sight, respectively.}
\end{deluxetable*}

\subsection{Outflow energy compared with gravitational energy} \label{sec:gravity}

We compared the outflow energy with the gravitational binding energy of the clump to investigate whether the outflows could potentially unbind the cloud.  Using the previously mentioned cloud mass and radius, the gravitational binding energy ($W=-GM_{cl}^2 /R_{cl}$) of the central region is $-2.5\times10^{46}$\ erg. \todo{Value}  The cloud mass is about 50 times greater than the outflow mass, while the gravitational binding energy is about 7 times greater than the outflow energy, assuming an inclination angle of $\xi=57.3\dg$ (see Table \ref{tab:mass}). \todo{Value}  In order to quantify whether outflows have sufficient energy to overcome gravity and disperse the surrounding gas, \citet{Nak14} utilized the non-dimensional parameter $\eta_{out}\equiv-2E_{out}/W$.  We found that $\eta_{out}=0.3$ \todo{Value}\ (assuming $\xi=57.3\dg$), which implies that outflow feedback is not sufficient to provide enough kinetic energy to unbind the entire cloud.

\subsection{Serpens South and NGC 1333: a comparison} \label{sec:compare}
Serpens South is the second region observed as part of a study of clustered star formation; in \citet{Plu13} we mapped the prototypical protostellar cluster NGC 1333 utilizing CARMA and FCRAO observations of CO $J=1-0$.  Given the consistency among the observations, here we compare Serpens South and the more evolved NGC 1333, and we present an empirical scenario of outflow-cluster interaction at different evolutionary stages. The comparisons that we discuss below are also summarized in Table \ref{tab:compare}.

\citet{Gut08} suggested that the high percentage of YSOs that are protostars is evidence of Serpens South's youth. Here we present a tally of YSOs in Serpens South using the most recent classifications following \citet{Eva07} and \citet{Mau11}, allowing a consistent comparison with NGC 1333.  We count Class 0, Class I, and flat-SED sources as protostellar sources, and these along with Class II/III sources comprise YSOs.  We count 45 protostellar sources and a total of 55 YSOs within the $\sim22$\ arcmin$^2$ (0.3 pc$^2$) region of Serpens South that we show outlined with a white rectangle in Figure \ref{fig:continuum}, of which 32 protostars and a total of 37 YSOs reside within the $\sim8$\ arcmin$^2$ (0.1 pc$^2$) inner-most cluster region outlined by the red rectangle in Figure \ref{fig:continuum}.  We compare this to the 26 protostellar sources out of a total of 55 YSOs in the 0.23 pc$^2$ region of NGC 1333 mapped by \citet{Plu13}.  

The protostellar number densities (number of protostars per unit area mapped) are 130 and 110 pc$^{-2}$, \todo{Value} respectively, for Serpens South and NGC 1333, showing that both of the regions we mapped are very densely populated with young members but about 20\% more so in Serpens South.  Further, in Serpens South the protostellar fraction (the ratio of protostellar sources to YSOs) is about 80\%, whereas in NGC 1333 this percentage is 50\%.  We also note that within the central region, outlined by the red rectangle in Figure \ref{fig:continuum}(b), the protostellar fraction of Serpens South rises to 90\%; the central region corresponding to the same physical area in NGC 1333 maintains a protostellar fraction of $\sim50$\%.  The higher percentage of young sources in Serpens South suggests that the majority of sources in this region have only recently begun to drive outflows that impact the surrounding cloud environment, compared with NGC 1333.

In NGC 1333, \citet{Plu13} identified outflows associated with 6 of 11 Class 0 sources and 2 of 11 Class I sources, but no outflows were associated with more evolved sources.  This corresponds to 8 of 26 protostellar sources driving identifiable outflows, and at least four candidate outflows had unidentifiable driving sources.  It is worth noting that NGC 1333 is closer (235 pc) than Serpens South (429 pc, see Section \ref{sec:SerpS}), and its outflow features are more clearly distinguishable such that we were able to associate outflows with their driving sources.  This is not the case for Serpens South, which is farther and has a much more complex web of interspersed outflows, as seen in Figure \ref{fig:carma_apex} and elsewhere, so our comparisons here are in terms of cumulative outflow activity.  In total, \citet{Plu13} measured 6\ \msun\ of gas associated with outflow lobes or outflow candidates in NGC 1333 within a region of $\sim0.23$\ pc$^2$, corresponding to $\sim26\ \msun$\ pc$^{-2}$.  In Serpens South,  we measure $\sim2.8\ \msun$\ of CO gas within a region of $\sim0.34$\ pc$^2$, or $\sim 10\ \msun$ pc$^{-2}$. \todo{Value}  Granted, a comparison of mass entrained by outflows relies on the assumption that the initial conditions of the cluster before star formation began are also comparable.

Spatially, outflow emission is very prevalent in both regions.  To quantify the fraction of the mapped area with outflow emission, we measured how many pixels (projected onto the plane of the sky) have \twco\ signal-to-noise ratio greater than three in at least three velocity channels (``outflow emission prevalence'' in Table \ref{tab:compare}).  In the Serpens South map, 80\% of the mapped region has emission in at least three velocity channels.  For comparison, we scaled emission in the NGC 1333 map from \citet{Plu13} to a distance of 429 pc and the same rms noise threshold as the Serpens South data.  In a region corresponding to the same physical area of the Serpens South map we measured emission (again in at least three velocity channels) spanning about 60\% of the mapped area on the plane of the sky.  The most simple explanation for this is that the number density of protostellar sources, assumed to drive outflows, is higher in the central-most region of Serpens South than in NGC 1333.  A more detailed study of individual outflows in Serpens South is needed to develop this argument.  Considering that NGC 1333 is a more evolved protostellar region, we expect some of the outflowing material to have escaped that region, whereas in Serpens South the material has had less time to travel beyond the central cluster that we mapped.  This interpretation of outflow emission prevalence compared between the two regions assumes that outflows are randomly oriented in both regions.  Alternatively, if large-scale filaments and gas kinematics cause preferential orientations of the outflows and/or their driving sources, then the spatial prevalence projected on the plane of the sky would depend on the inclination angle of the majority of outflows.  Since Serpens South is known to have a concentration of YSOs along a dense filamentary structure \citep{Gut08}, then this structure potentially affects the spatial distribution of outflow emission.

Two previously mentioned ratios, $r_L$ and $\eta_{out}$, quantify the relative contributions of energies from outflows and turbulence, and from outflows and gravity (respectively) in these regions.  For the slightly more evolved region NGC 1333, these parameters are $r_L\sim6-24$ and $\eta_{out}=0.2-0.7$ (for $\xi=0-57.3\dg$), whereas in Serpens South we found these values to be $r_L\sim0.6-3$ and $\eta_{out}=0.1-0.3$ (for $\xi=0-57.3\dg$), respectively, which are about 10-50\% the values for NGC 1333. \todo{Value}  The outflows in both regions contribute sufficiently to drive turbulence, although to a greater extent in NGC 1333, and in NGC 1333 the outflows may nearly counter the gravitational energy of the clump.  It follows that as a cluster evolves from the very young stage of Serpens South to the slightly more evolved stage of NGC 1333, the outflows sustain turbulence in a similar proportion, but destruction via outflows is more imminent as a region evolves.

Finally, we explore an evolutionary scenario in which Serpens South may come to resemble NGC 1333.  First, we calculated the relative protostellar fraction within Serpens South as it progresses from its current, young age \citep[$\sim2\pm1\times10^5$\ yr,][]{Fri13} to the slightly more evolved state of NGC 1333 over its expected age of $\sim1-2$\ Myr \citep{Lad96}.  We assumed lifetimes of 0.10, 0.34, and 0.35 Myr spent as Class 0, Class I, and Flat SED classes, respectively \citep{Eva09}, and a uniform distribution of ages within each of these classes.  In the simplest (unlikely) case, if no additional protostars form, then the protostellar fraction will reach that of NGC 1333 ($\sim50\%$) in 0.4 Myr.  In order to account for the ongoing formation of additional, young protostars, then incorporating also the star formation rate (SFR) of 90 \msun\ Myr$^{-1}$\ \citep{Gut08}, and an average YSO mass of 0.5 \msun\ \citep[c.f.][]{Eva09}, then we add one more protostar every 0.006 Myr, and in that case the protostellar fraction reaches 50\% in 1.3 Myr.  However, the SFR in the more evolved region NGC 1333 is about half of that in Serpens South, suggesting that the SFR likely decreases as a cluster evolves.  If instead we assume the SFR over the entire time to be 45 \msun\ Myr$^{-1}$\  \citep[as for NGC 1333,][]{Lad96}, then the protostellar fraction reaches 50\% in 0.9 Myr.  Additional work should be done to determine how the SFR changes over time for a cluster, but even based on these simple calculations using representative values for the SFR, the protostellar fraction of Serpens South will likely decrease to be comparable to that of NGC 1333 in $\sim1$\ Myr.

Also, as Serpens South evolves to the age of NGC 1333 during the next $\sim1$\ Myr, we expect mass to escape from the cloud, possibly also related to the respective power-law slopes of the mass spectra in Figure \ref{fig:mass_spectra}.  We calculated the escape velocity of Serpens South to be $v_{esc}=3.4$\ \kms, which is higher than that of NGC 1333 where $v_{esc}=2.2$\ \kms\ \citep{Arc10}, but nonetheless most of the outflow mass we calculate in both regions pertains to gas with outflow velocities greater than the escape velocity.  Assuming ages of Serpens South and NGC 1333 as above, and assuming a constant outflow rate, then the current mass loss rate of $70-120\ \msun$\ Myr$^{-1}$ implies that $\sim100\ \msun$ \todo{Value} may be lost before Serpens South reaches the evolutionary stage of NGC 1333.  At that time, with a cloud mass that is about half of what we assumed here for the calculations of gravitational energy of the clump, the value of $\eta_{out}$ in Serpens South will approximate that of NGC 1333.  If cumulative outflow energy decreases with time, then Serpens South would come to resemble NGC 1333 sooner.  Although this is a simple order of magnitude calculation with admittedly simple assumptions, we consider it valuable to place Serpens South and NGC 1333 within a relative evolutionary scenario.  More thorough treatment of the assumptions we made here, including accretion onto the filament \citep{Kir13} and varying (and overall decreasing) mass ejection rate over time, as well as further detailed analysis of additional clusters spanning a range of evolutionary stages will constrain this scenario.

\section{Summary} \label{sec:summary}

Here we assess the activity of outflows in the young protostellar region Serpens South, specifically focusing on the central 0.3 pc$^2$ where we find an intricate web of outflow activity.  We present observations of CO $J=1-0$\ for which interferometer (CARMA) and single dish (IRAM 30 m telescope) observations were combined to probe a range of spatial scales from approximately 0.01 pc (or 2000 AU) to 0.8 pc (at a distance of 429 pc).  These provide the highest-resolution observations of molecular outflows in this region to date, while subsequently retaining extended cloud-scale emission.  

Utilizing the CO $J=1-0$\ data from CARMA and the IRAM 30 m telescope, in addition to CO $J=3-2$\ data from APEX and CSO, we determined an opacity correction factor for the main beam temperature of the \twco\ observations as well as solved for varying excitation temperature throughout the region; these methods are described in Appendix \ref{sec:opacity} and \ref{sec:tex}.  These observations and methods were necessary to obtain accurate estimates of outflow mass, momentum and energy.  The results can be summarized as the following:

\begin{enumerate}
\item Within the region we mapped, we measured the total outflow mass to be $M_{out}=2.8\ \msun$, outflow momentum to be $P_{out}=13.6\ \msun\ \kms$, and energy to be $E_{out}\sim1.0\times10^{45}$\ erg (uncorrected for assumed inclination angles of outflows). \todo{Value} Several methods using different assumed excitation temperatures resulted in masses, momenta and energies that varied by factors of a few.  

\item The outflow mass spectra were fit with power laws of $3.5\pm0.1$\ and $3.3\pm0.1$\todo{Value}\ for the blueredshifted and redshifted emission, respectively, for velocity channels with $3 < |v_{out}|< 15\ \kms$.  We found steeper slopes for the outflows in the more evolved region NGC 1333 than for Serpens South, and we found no break for either region, having properly accounted for optical depth effects and recovered the ``hidden'' mass at low outflow velocities.  Likely a shallower (steeper) slope is a sign of the earlier (later) evolutionary state of a cluster.

\item We quantified the relative contributions of outflows compared with turbulence and gravity in this region, comparing the outflow energy input rate to the turbulence energy dissipation rate and the outflow energy to the gravitational energy, as well as outflow momentum force to the turbulence momentum dissipation rate.  We found that outflows contribute enough momentum and energy to maintain turbulence, but outflow feedback plays a minor role compared with the gravitational potential energy of the clump.  These mass, momentum, and energy diagnostics are consistent with the quasi-static star formation model.  Additionally, based on the cumulative mass loss rate, we estimated a mass accretion rate that is comparable to the lower limit infall rate onto the filament.

\item In Serpens South, the bulk of the outflow emission appears to be emanating from sources located within a central 0.1 pc$^2$ region of high protostellar density.  Our 2.7 mm continuum observations with CARMA show seven distinct continuum emission sources that are considered protostellar candidates, some of which correspond to the $\sim45$ \textit{Spitzer} protostars in this region.  The high resolution of our observations reveals the extended morphologies of CARMA-3a/b and CARMA-6/7, and we measure projected distance separations for each proposed binary (or multiple) system of about $3000$\ and $6500$\ AU, respectively.  

\item In a census of YSOs in Serpens South compared with NGC 1333, the central hub of the Serpens South cluster contains a relatively high protostellar fraction and protostellar number density, and we compare outflow energetics and their influence on the respective clouds.

\end{enumerate}

Considering these outflow diagnostics and the YSO census in Serpens South, we propose an empirical evolutionary scenario for which Serpens South reveals an early stage of star formation, and this complex region will likely evolve to resemble a protostellar region like NGC 1333 within approximately 1 Myr.  Even higher resolution continuum emission observations are needed to reveal several sources as likely multiple-member systems, and higher sensitivity observations will  reveal weaker sources, refining our view of the young sources that drive outflows in this region.  Similarly, higher resolution molecular outflow observations, of which ALMA is capable (A.L. Plunkett et al. in prep), will allow us to associate specific outflows with their driving sources, and better characterize individual outflows as was done for NGC 1333 by \citet{Plu13}.

\acknowledgments

We thank the anonymous referee for providing careful comments to improve the quality of the paper.  We would like to thank Manuel Fern{\'a}ndez-L{\'o}pez and Jenny Hatchell for specific details and discussion of this region.  A. L. P. acknowledges the Yale Graduate Writing Center Peer Review Group for constructive comments about this manuscript throughout the writing process.  A. L. P. is supported by the National Science Foundation Graduate Research Fellowship under Grant No. DGE-1122492, and previously by the US Student Program of Fulbright Chile.  H. G. A. receives funding from the NSF under grant AST-0845619.  M. M. D. acknowledges support from the Submillimeter Array through an SMA postdoctoral fellowship.  D. M. and G. G. acknowledge support from CONICYT project PFB-06.  This publication is based on data acquired with the Combined Array for Research in Millimeter-wave Astronomy (CARMA), IRAM 30 m telescope, Atacama Pathfinder Experiment (APEX), and Caltech Submillimeter Observatory (CSO).  Support for CARMA construction was derived from the Gordon and Betty Moore Foundation, the Kenneth T. and Eileen L. Norris Foundation, the James S. McDonnell Foundation, the Associates of the California Institute of Technology, the University of Chicago, the states of California, Illinois, and Maryland, and the National Science Foundation. Ongoing CARMA development and operations are supported by the National Science Foundation under a cooperative agreement, and by the CARMA partner universities.  The 30 m telescope, operated by the Institut de RadioAstronomie Millimétrique (IRAM), is funded by a partnership of INSU/CNRS (France), MPG (Germany), and IGN (Spain).  APEX is a collaboration between the Max-Planck-Institut fur Radioastronomie, the European Southern Observatory, and the Onsala Space Observatory.  The CSO is operated by the California Institute of Technology under cooperative agreement with the National Science Foundation (AST-0838261).  The research presented here made use of Astropy, a community-developed core Python package for Astronomy (Astropy Collaboration, 2013)\nocite{astropy}.

{\it Facilities:} \facility{CARMA}, \facility{IRAM}, \facility{APEX}, \facility{CSO}.

\appendix

\section{Correcting for opacity} \label{sec:opacity}
In order to accurately measure masses and energetics of outflows, \citet{Dun14} argue that careful attention must be paid to account for the effects of line opacity and excitation temperature variations (see Appendix \ref{sec:tex}), among other factors.  We follow the method presented by \citet[][and references therein]{Dun14} to correct for opacity, given that low-$J$\ transitions of \twco\ are typically optically thick at cloud systemic velocities, and this likely needs to be accounted for in the line-wings as well.  Here, opacity is determined for each velocity channel, independently for the $J=1-0$\ and $J=3-2$\ energy level transitions.  For the $J=1-0$\ transition, we used the combined CARMA+IRAM maps, and the \thco\ data cube was regridded to have the same pixel size and channel width as the \twco\ data cube.  Units of the CARMA+IRAM maps were converted from Jy beam$^{-1}$\ to Kelvin by dividing by 3.78 and 4.13 for \twco\ and \thco, respectively.   For the $J=3-2$\ transition, we used data from APEX and CSO for \twco\ and \thco, respectively, after convolving the APEX data cubes to have the same beamsize (35\as) as the CSO observations, and we smoothed the CSO spectra to have the same velocity resolution (0.2 \kms) as the APEX observations (see Sections \ref{sec:APEX} and \ref{sec:CSO}).   

\begin{figure*}[!ht]
\includegraphics[width=\linewidth,angle=0]{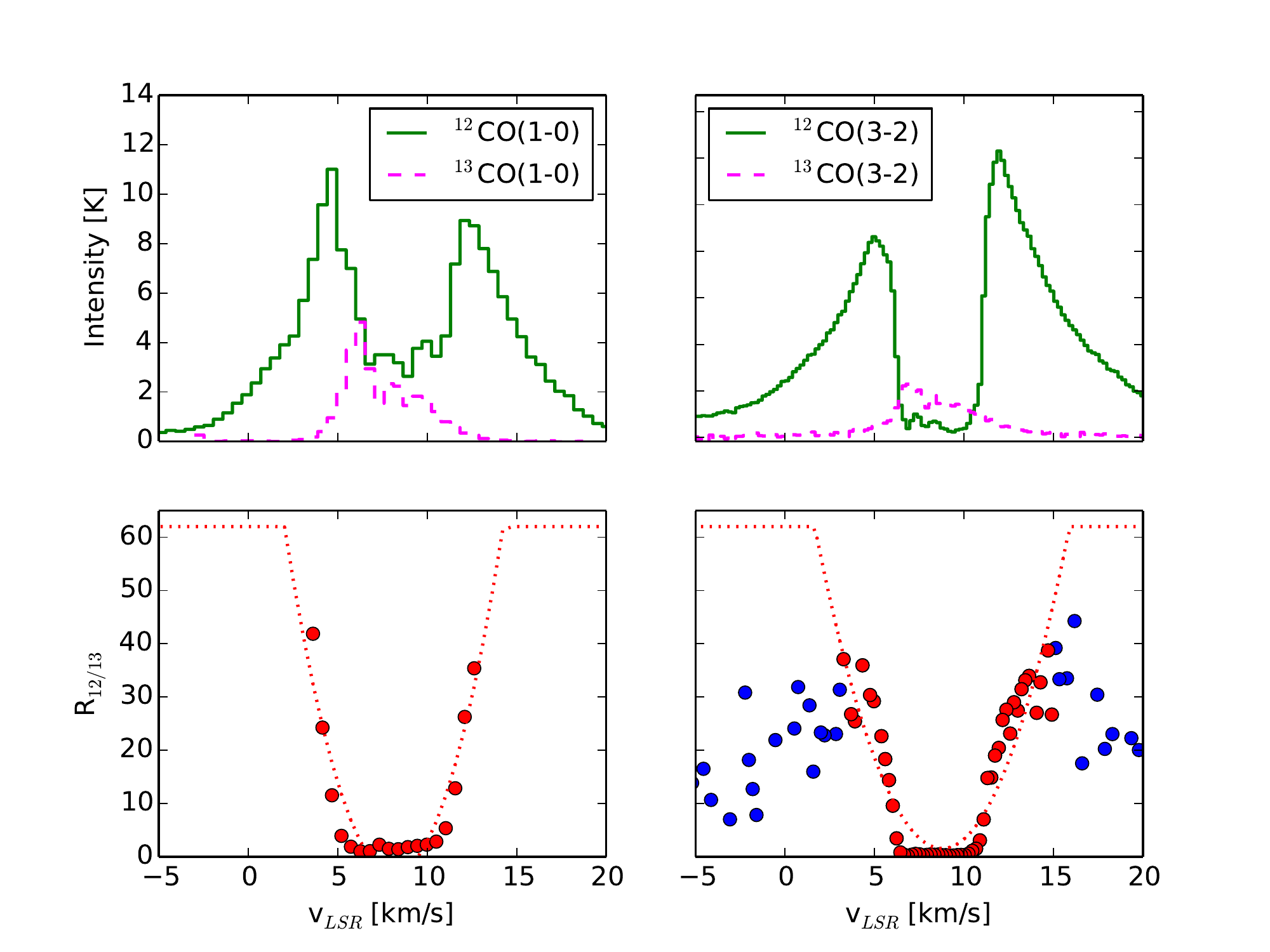}
\caption{Top row shows \twco\ (solid green line) and \thco\ (dashed magenta line) spectra for the $J=1-0$\ (left) and $J=3-2$\ (right) transitions, averaging over the regions specified in the text (see Appendix \ref{sec:opacity}).  Bottom row shows the ratio of these spectra for both transitions, $R_{12/13}=I_{12}/I_{13}$.  A parabola, shown with a red dotted line, was fit to the points also shown in red for each transition, and the equations for these parabolas are given in Equations (\ref{eq:para10}) and (\ref{eq:para32}), truncated at a value of 62. Channels with \thco\ emission below $3\sigma$ and/or $R_{12/13}>62$ are excluded from the plot.   }  
\label{fig:opacityfits}
\end{figure*}

\begin{figure}[!ht]
\includegraphics[width=0.5\linewidth,angle=0]{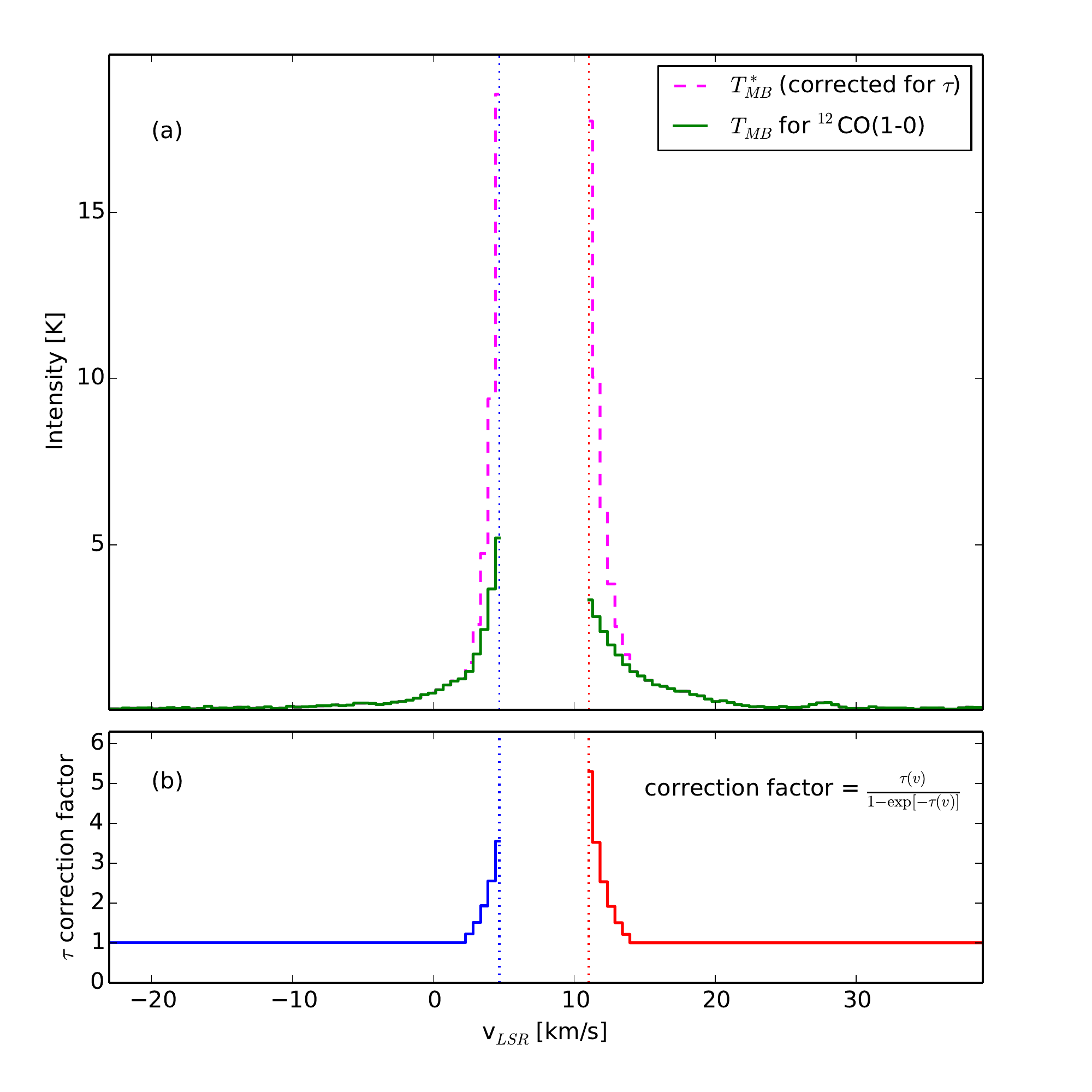}  
\caption{(a) Average \twco\ $J=1-0$\ spectrum within the central $\sim16$\ arcmin$^2$ region of the map.  The green (solid) line shows the \twco\ emission without any correction for opacity, and the magenta (dashed) line shows the spectrum after correcting for optical depth of the line, according to Equation \ref{eq:tmbcorr}.  Vertical blue and red dotted lines mark the velocities $v_{cloud}\pm3$\ \kms, and we only consider velocity channels outside of this range to be mostly unaffected by cloud emission. (b) Factor used to correct \twco\ main beam temperature for optical depth (see equation \ref{eq:tmbcorr}). }  
\label{fig:opacitycorrection}
\end{figure}

Figure \ref{fig:opacityfits} shows the channel-by-channel intensity of \twco\ and \thco. For the $J=1-0$\ maps, we averaged within a box $60\times60\as$ centered at R.A.$=18^h30^m03^s$, decl.$=-02\dg02\arcmin57\arcsec$, where a high prevalence of outflow emission is observed.  In the case of the $J=3-2$\ maps, we averaged the region of each map corresponding to the two CSO pointings nearest to the center (see Figure \ref{fig:cover}).   Independently for $J=3-2$\ and $J=1-0$, we fit a parabolic function to the ratio $R_{12/13}(v_i)=\langle I_{12}(v_i)\rangle/\langle I_{13}(v_i)\rangle$, where $\langle I_{12}(v_i)\rangle$ and $\langle I_{13}(v_i)\rangle$ are averaged intensities for each velocity channel $v_i$.  The parabolic fits are described with the following equations: 
\begin{eqnarray} 
J=1-0: && R_{12/13}(v_i) = (1.75\pm0.03) (v_i-8.14\pm0.02)^2 + (-3.2\pm3.2) \label{eq:para10}\\ 
J=3-2: && R_{12/13}(v_i) = (1.19\pm0.01)(v_i-8.78\pm0.01)^2 + (1.6\pm1.6). \label{eq:para32}
\end{eqnarray}
We fit the function to the velocity ranges (selected by eye to reasonably fit a parabolic function) indicated with red points in Figure \ref{fig:opacityfits}, and we truncated each function with a maximum value of 62 at the high-velocity edges of the spectra. Also, we do not constrain the fits to reach a minimum at the cloud velocity, as can be seen from the equations above which both have minima at velocities greater than 8 \kms.  This asymmetrical opacity correction may be an effect of averaging over a region that is comprised of several outflows with distinct orientations and inhomogeneous initial cloud gas distribution, and therefore the averaged outflow emission is inherently asymmetric.  Using these functions, we numerically determined the velocity-dependent optical depth $\tau_{12}$ for both transitions, following the relation:

\begin{equation}
R_{12/13}= \frac{[\twco]}{[\thco]} \left(\frac{1-\exp[-\tau_{12}]}{\tau_{12}}\right),
\end{equation}
assuming that the \thco\ is optically thin.  We then correct the observed main-beam temperature, $T_{mb,12}$, for the \twco\ velocity-dependent opacity, resulting in the corrected main-beam temperature, $T_{mb,12}^*$, given by:
\begin{equation} \label{eq:tmbcorr}
T_{mb,12}^*=T_{mb,12} \frac{\tau_{12}}{1-\exp(-\tau_{12})}.
\end{equation}
In Figure \ref{fig:opacitycorrection} we show the averaged \twco\ spectrum versus velocity before and after making this opacity correction, as well as the factor that relates observed main-beam temperature and corrected main-beam temperature.  The correction factor shown in Figure \ref{fig:opacitycorrection}(b) reaches $\sim5$ for velocity channels with $|v_{LSR}-v_{cloud}|\sim3\ \kms$\, and reduces to 1 for velocities with $|v_{LSR}-v_{cloud}|\gtrsim 6\ \kms$.

\begin{figure}[!ht]
\includegraphics[width=0.5\linewidth,angle=0]{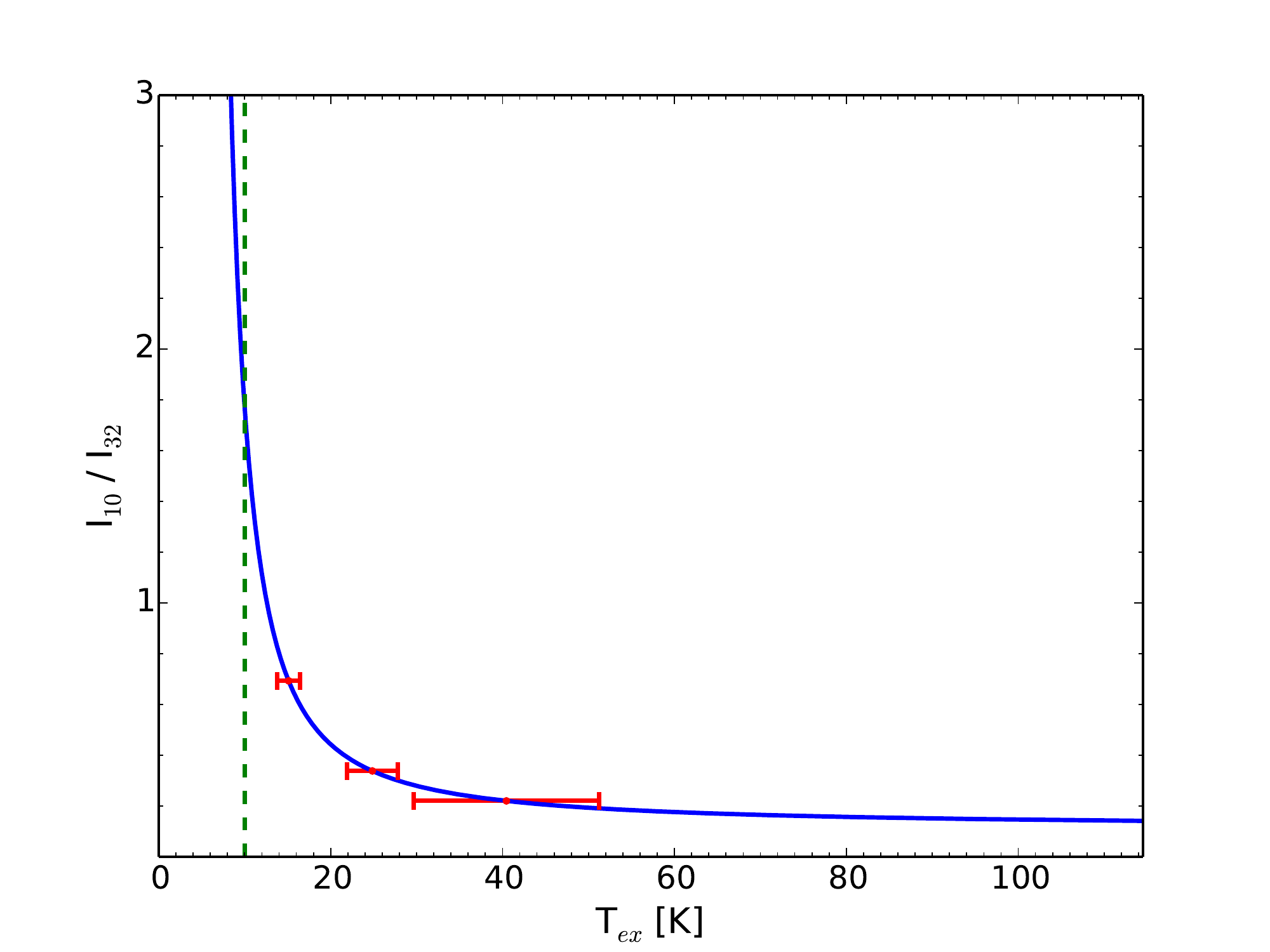}
\caption{Ratio of intensities I$_{10}$/I$_{32}$ expected as a function of excitation temperature.  See Appendix \ref{sec:tex}, where we describe how we account for varying \tex\ in our map.   The vertical dashed green line marks the imposed minimum at $\tex=10$ K.  Several sample error bars show uncertainty in \tex, based on the rms of the \twco\ $J=3-2$\ and $J=1-0$\ single-dish maps.  }  
\label{fig:tex_vs_R}
\end{figure}

\section{Excitation temperature} \label{sec:tex}
In addition to opacity correction, excitation temperature (\tex) variations have an effect on measuring outflow characteristics.  Specifically, \citet{Dun14} show that outflow mass is uncertain by a factor of 0.7-3 depending on assumed \tex\ in the range of 10-200 K.  Excitation temperature is generally assumed to be within the range of about 10 - 50 K and perhaps higher \citep[see discussion in][]{Dun14} in protostellar regions such as Serpens South.  

Given the range of plausible values, rather than adopt a single value for the entire region, we opted to obtain the value on a pixel-by-pixel + channel-by-channel basis using the \twco\ $J=1-0$\ and $J=3-2$\ data.  We used observations from IRAM (\twco\ $J=1-0$) and APEX (\twco\ $J=3-2$) to solve for H$_2$ column density, $N_{H_2}$, as a function of excitation temperature according to equations \ref{eq:NH2} and \ref{eq:f}.  We did so independently using single dish observations of these two transitions, and then set the two equal to solve for \tex\ in each pixel and velocity channel.  In other words, excitation temperature can be determined from the ratio of $J=1-0$\ to $J=3-2$\ intensities, $I_{10}/I_{32}$, as we show in Figure \ref{fig:tex_vs_R} and according to the following equation:
\begin{eqnarray} \label{eqn:tex}
T_{ex}=\frac{(E_{J+1,10}-E_{J+1,32})}{k}\left(\ln\left[ \frac{\nu_{10}}{\nu_{32}}\frac{(J_{10}+1)}{(J_{32}+1)} \frac{I_{32}}{I_{10}}\right]\right)^{-1},
\end{eqnarray}
where we used the opacity-corrected main-beam temperature ($T_{mb,12}^*$, see equation \ref{eq:tmbcorr}) for each \twco\ transition as $I_{10}$\ and $I_{32}$, respectively.  We used the following values for the $J=1-0$\ transition: $J_{10}=0$, $E_{J+1,10}=7.64\times10^{-16}$ erg.  Similarly, for the $J=3-2$\ transition we use the following values: $J_{32}=2$, $E_{J+1,32}= 4.58\times10^{-15}$ erg, and $E_{J+1,32}/k=33.19$ K.

Solving for \tex\ in pixels where $I_{10}$ and $I_{32}$ are above 4$\sigma$, we measured excitation temperatures to be in the range $10$\ K $\lesssim \tex \lesssim 100$\ K.  Where we could not reliably determine \tex\ for a given pixel, if we could do so for neighboring pixels, we took the averaged $\tex$ in the surrounding box of $5\times5$ pixels and velocity channels within $v_{out}=\pm 0.5\ \kms$.  Finally, we set a lower limit of $\tex=10$ K for any pixels that did not meet the 4$\sigma$ criterion, as well as pixels beyond the boundary of the $\sim 5^\prime \times 5 ^\prime$ APEX map (see Figure \ref{fig:cover}). 

\begin{figure}[!ht]
\includegraphics[width=0.5\linewidth,angle=0]{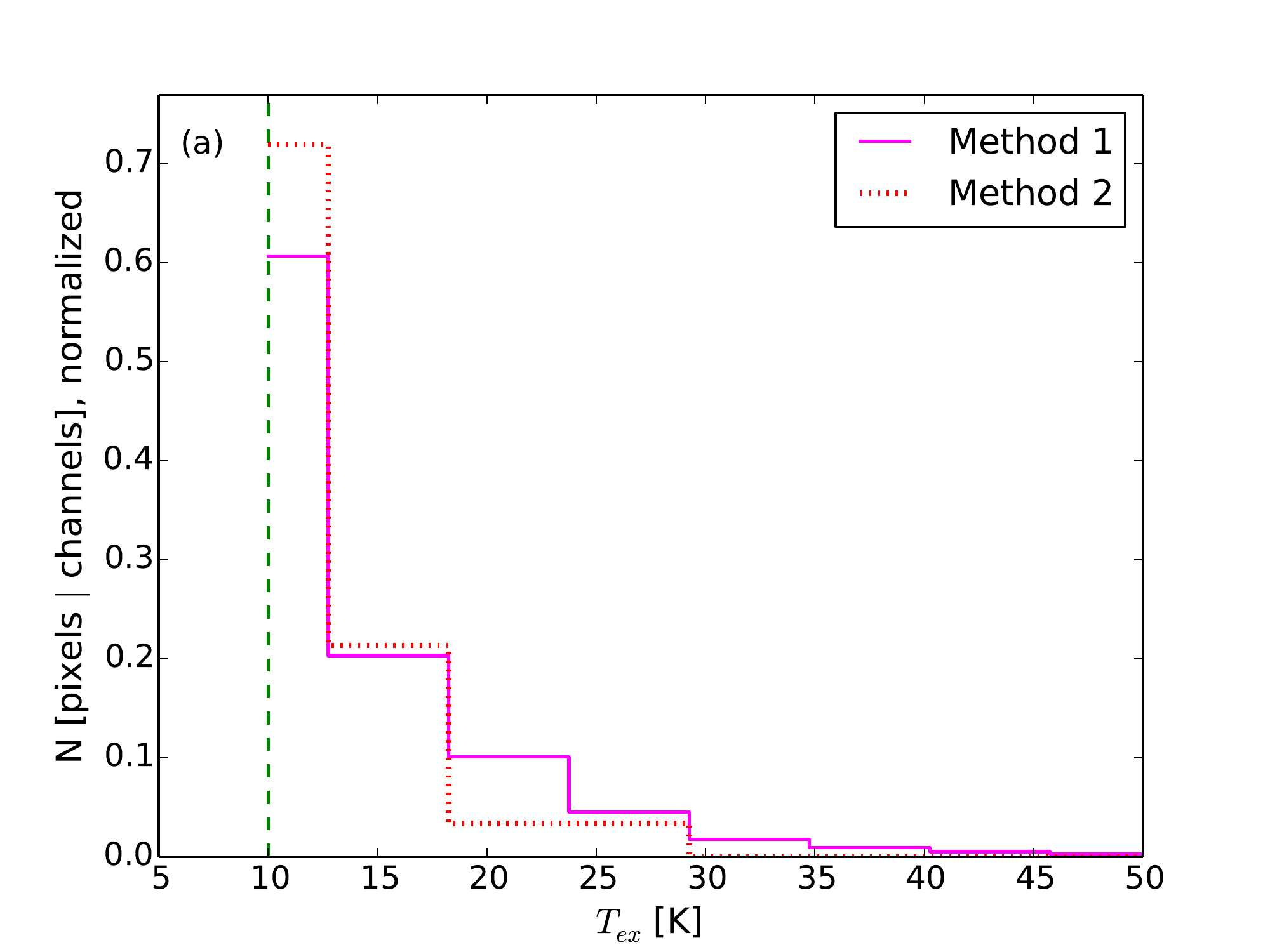}
\includegraphics[width=0.5\linewidth,angle=0]{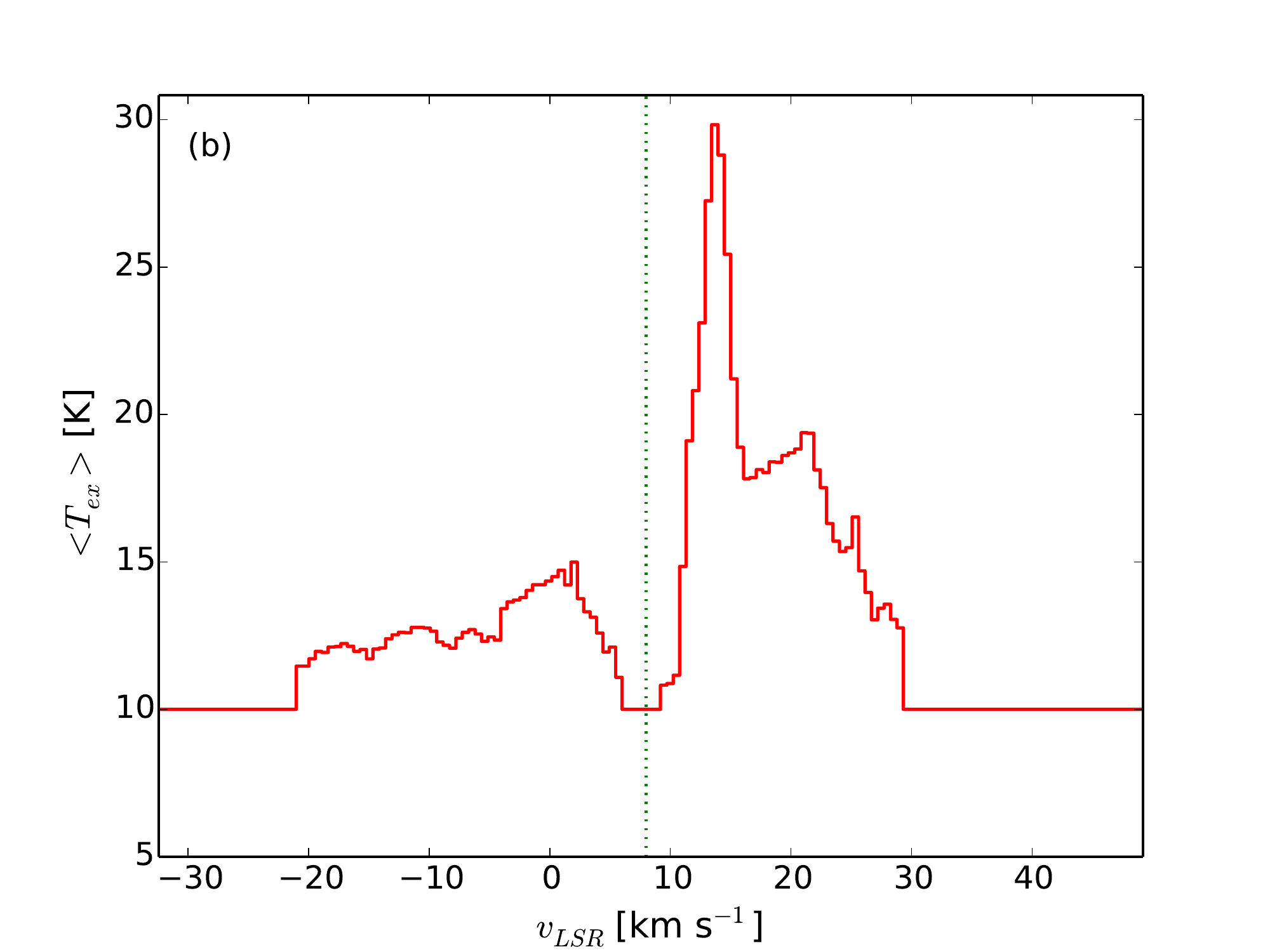}
\caption{(a) Normalized distribution of excitation temperatures determined pixel-by-pixel and channel-by-channel (method 1, magenta solid line) and spatially averaging \tex\ channel-by-channel (method 2, red dotted line). The vertical dashed green line marks the imposed minimum at $\tex=10$ K. (b) Excitation temperature averaged for each velocity channel (method 2). The vertical dotted green line corresponds to the central cloud velocity of \vcloud.}  
\label{fig:texhist}
\end{figure}

\begin{figure}[!ht]
\includegraphics[width=.8\linewidth]{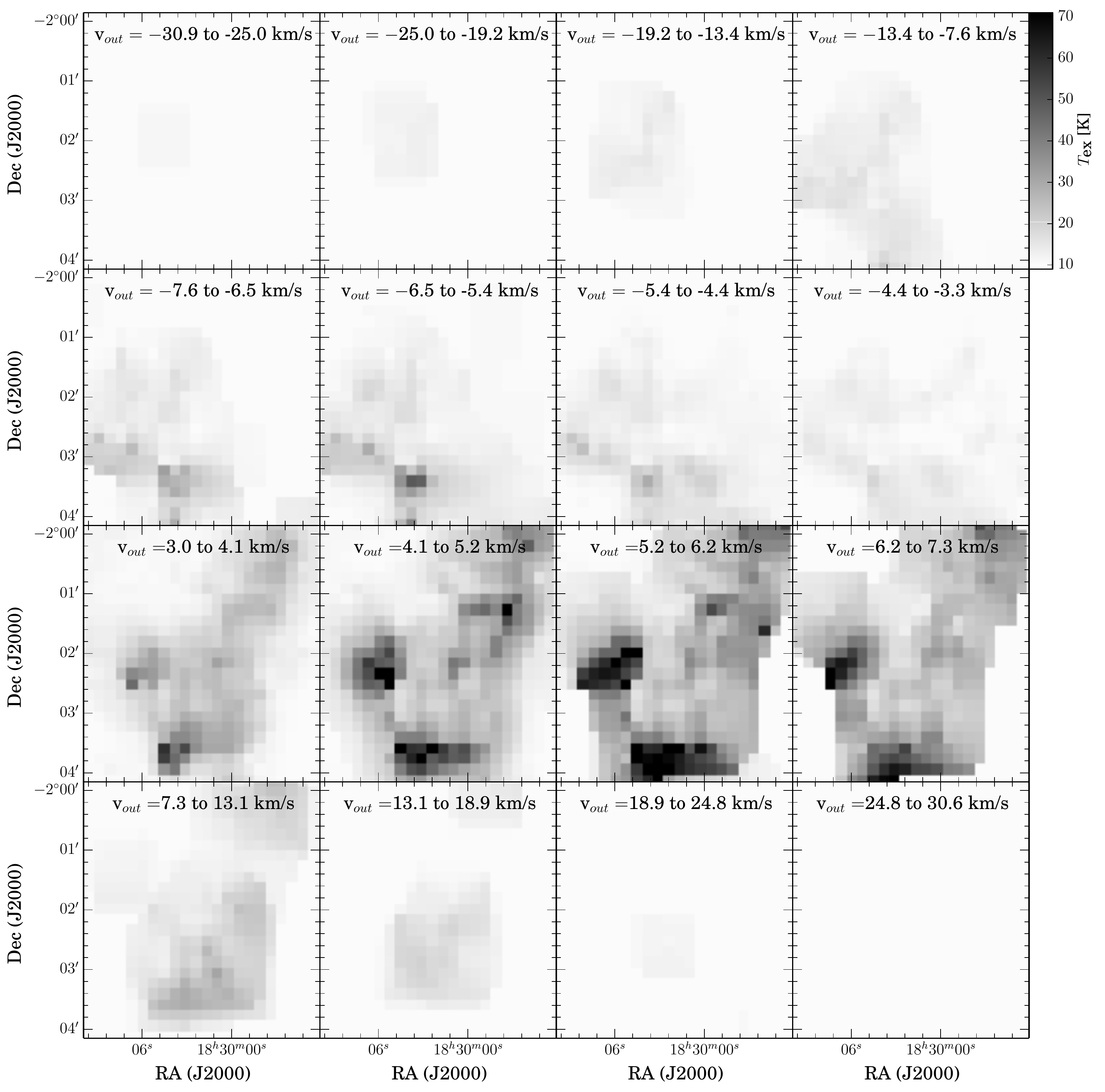}
\caption{Maps of $\tex$ for each pixel position, averaged at each pixel position within the channel intervals shown for each panel (these are the same intervals used in Figure \ref{fig:channel2}).  Color scale bar is shown in the upper right, and chosen to truncate at $\tex=70$\ K to show color contrast in the maps.  We measured excitation temperatures to be in the range $10$\ K $\lesssim \tex \lesssim 100$\ K.}  
\label{fig:channel_tex}
\end{figure}

We then regridded the excitation temperature data to the grid of the CARMA+FCRAO map so that each pixel in the combined interferometer and single dish map, used for determining mass, had a corresponding excitation temperature.  The map of excitation temperature is shown in Figure \ref{fig:channel_tex}.  The pixel-by-pixel + channel-by-channel method described here, which we call method 1, results in an average excitation temperature of $\overline{\tex}(x,y,v)=17 \pm 8$\ K.  For comparison with another protostellar region, in NGC 1333 \citep{Plu13} we performed the same velocity-dependent pixel-by-pixel approach and found $\overline{\tex}(x,y,v)=20\pm4$ K.  In Serpens South, we also calculated $\tex(v)$ channel-by-channel (method 2), where $\tex(v)$ is the excitation temperature obtained for each channel by spatially averaging over the region (again, using the opacity-corrected main-beam temperature) and adopted for all pixels in that channel.  By method 2, we found $\overline{\tex}(v)=15 \pm 4$\ K.  Finally, we present three methods (3-5) that assumed constant values of  $\tex=10,\ 30,\ 50$ K to investigate this effect on mass.  The distributions of excitation temperature values in the pixel-by-pixel + channel-by-channel (method 1) and channel-by-channel (method 2) approaches are shown in Figure \ref{fig:texhist}. 

We calculated the error in \tex\ by propagating the rms errors of $I_{10}$ and $I_{32}$ in Equation \ref{eqn:tex}, and we show the errors (for method 1) incurred for several representative excitation temperatures in Figure \ref{fig:tex_vs_R}.  The error increases for higher \tex\ pixels; for example the errors in \tex\ are about $\pm 1$, $\pm 10$, and $\pm 30$ K  for \tex=10, 30, and 50 K, respectively. \todo{Value}

Mass, momentum and energy that we mention throughout the text (unless explicitly noted otherwise) are those determined using $\tex(x,y,v)$\ of method 1, taking into account the spatial- and velocity-dependence of excitation temperature, and in Table \ref{tab:mass} we present these values for all \tex\ methods.  Our five methods of \tex\ result in a factor of three difference for each outflow property (mass, momentum, and energy), with a constant \tex=50 K resulting in the highest values and constant \tex=10 K resulting in the lowest values (see Table \ref{tab:mass}).  Considering that the velocity-dependent pixel-by-pixel method shows that \tex\ ranges from 10 to 100 K within the region, then assuming a constant \tex\ will likely under-estimate (over-estimate) the mass in regions where the actual \tex\ is higher (lower) than the assumed constant \tex.  This is consistent with the work of \citet{Dun14}, who showed that outflow masses determined using \twco\ $J=2-1$\ and \twco $J=3-2$\ varied by factors of $0.7-3$ times depending on assumed excitation temperature in the range \tex=10-200 K, relative to the mass determined using their assumed \tex=50 K.  

Ideally, we would use several other energy level transitions in addition to the two that we used here to more accurately constrain the excitation temperatures as a function of position and/or velocity.  Observations of high-$J$ CO lines in other regions have revealed warmer excitation temperatures in the range $50<T<200$ K associated with higher velocity gas \citep{Van09a,Van09b,Yil13}.  Likely the warmer gas is due to recent shocks, while colder (and slower) gas is entrained after being shocked in the past.  Based on higher-$J$\ transitions of CO, \citet{Yil13} measured higher excitation temperatures (\tex $>50$\ K) than the mean values measured in our study, and they showed \tex\ to increase with outflow velocity.  Hence, the high-$J$\ CO emission is important for probing the warmer and/or denser gas components, which are associated with broader line wings.  

Our lower-$J$\ CO data correspond to the cooler, less dense gas components with narrower line wings; we show that excitation temperatures peak at an outflow velocity of about 5-10 \kms, and decrease at higher velocities. Since \citet{Yil13} claim that the broad and narrow components contribute roughly equally, and we are only probing the cooler, narrow component, then only including the $J=1-0$\ and $J=3-2$\ transitions in our calculations likely underestimates the \tex\ measurement and subsequent mass, momentum, and energy determinations.  It follows that the diagnostics that we consider to assess the contribution of outflows to the energetics of the cloud, including $r_L$\ and $\eta_{out}$\ (see Sections \ref{sec:turbulence}-\ref{sec:gravity}), should also be considered lower limits.  Hence, the contribution of outflows may be more significant than we report here, reinforcing the main finding that outflows are important to the kinematics of this young cluster.

\bibliographystyle{apj}

\end{document}